\documentclass[apj]{emulateapj}

\newcommand{\etal}{{et al.~}}
\newcommand{\msunh}{\>h^{-1}\rm M_\odot}

\newcommand{\Msun}{\>{\rm M_{\odot}}}
\newcommand{\Lsun}{\>{\rm L_{\odot}}}
\newcommand{\mpch}{\>h^{-1}{\rm {Mpc}}}
\newcommand{\kms}{\>{\rm km}\,{\rm s}^{-1}}
\newcommand{\kmsmpc}{\>{\rm km}\,{\rm s}^{-1}\,{\rm Mpc}^{-1}}
\newcommand{\Qmag}{\>^{0.1}{\rm M}_Q-5\log h}
\newcommand{\rmag}{\>^{0.1}{\rm M}_r-5\log h}
\newcommand{\rrmag}{\>^{0.0}{\rm M}_r-5\log h}

\newcommand{\rmaglim}{\>^{0.1}{\rm M}_{r,{\rm lim}}-5\log h}
\newcommand{\calC}{{\cal C}}
\newcommand{\rmd}{{\rm d}}

\shorttitle{Galaxy Groups in the SDSS DR4}
\shortauthors{Yang et al.}

\begin{document}
            

\title{Galaxy Groups in the SDSS DR4: I. 
       The Catalogue and Basic Properties}    
    
\author{Xiaohu Yang\altaffilmark{1,4}, H.J. Mo \altaffilmark{2},
        Frank C. van den Bosch\altaffilmark{3}, Anna Pasquali\altaffilmark{3},
        Cheng Li\altaffilmark{1,4}, Marco Barden\altaffilmark{3,5}}

      \altaffiltext{1}{Shanghai Astronomical Observatory,
        the Partner Group of MPA, Nandan Road 80, Shanghai 200030, China; 
        E-mail: xhyang@shao.ac.cn}
      \altaffiltext{2}{Department of Astronomy, University of Massachusetts,
        Amherst MA 01003-9305}
      \altaffiltext{3} {Max-Planck-Institute for Astronomy, K\"onigstuhl 17, 
        D-69117 Heidelberg, Germany }
      \altaffiltext{4}{Joint Institute for Galaxy and Cosmology (JOINGC) of
        Shanghai Astronomical Observatory and University of Science and
        Technology of China}
      \altaffiltext{5}{Institut f\"ur Astrophysik, Leopold-Franzens
        Universit\"at Innsbruck, Technikerstrasse 25, A-6020 Innsbruck,
        Austria}


\begin{abstract}
  We use a modified version of the halo-based group finder developed by Yang
  et al. to select galaxy groups from the Sloan Digital Sky Survey (SDSS DR4).
  In the first step, a combination of two methods is used to identify the
  centers of potential groups and to estimate their characteristic luminosity.
  Using an iterative approach, the adaptive group finder then uses the average
  mass-to-light ratios of groups, obtained from the previous iteration, to
  assign a tentative mass to each group.  This mass is then used to estimate
  the size and velocity dispersion of the underlying halo that hosts the
  group, which in turn is used to determine group membership in redshift
  space. Finally, each individual group is assigned two different halo masses:
  one based on its characteristic luminosity, and the other based on its
  characteristic stellar mass.  Applying the group finder to the SDSS DR4, we
  obtain $301237$ groups in a broad dynamic range, including systems of
  isolated galaxies.  We use detailed mock galaxy catalogues constructed for
  the SDSS DR4 to test the performance of our group finder in terms of
  completeness of true members, contamination by interlopers, and accuracy of
  the assigned masses.  This paper is the first in a series and focuses on the
  selection procedure, tests of the reliability of the group finder, and the
  basic properties of the group catalogue (e.g.  the mass-to-light ratios, the
  halo mass to stellar mass ratios, etc.). The group catalogues including the
  membership of the groups are available at these
  links\footnote{http://gax.shao.ac.cn/data/Group.html \\
    http://www.astro.umass.edu/$^\sim$xhyang/Group.html}.
\end{abstract}




\section{Introduction}

Galaxies are thought  to form and reside in  extended cold dark matter
haloes. One of the ultimate challenges in astrophysics is therefore to
obtain  a  detailed  understanding  of  how  galaxies  with  different
physical properties occupy dark matter haloes of different mass.  This
relationship  not   only  conveys  important   information  about  how
different galaxies  form and evolve  in different dark  matter haloes,
but it also provides the  necessary basis for translating the observed
distribution of  galaxies into the large-scale  distribution of matter
throughout the Universe.

Theoretically,  the  relationship  between  galaxies and  dark  matter
haloes  can  be  studied  using  numerical  simulations  (e.g.,  Katz,
Weinberg \& Hernquist 1996; Pearce \etal 2000; Springel 2005; Springel
\etal  2005) or  semi-analytical models  (e.g.  White  \&  Frenk 1991;
Kauffmann  \etal 1993, 2004;  Somerville \&  Primack 1999;  Cole \etal
2000; van den Bosch 2002; Kang \etal 2005; Croton \etal 2006). Both of
these  techniques try  to model  the  process of  galaxy formation  ab
initio.   However, since  our  understanding of  the various  physical
processes involved is still relatively poor, the relations between the
properties of galaxies and their dark matter haloes predicted by these
simulations and semi-analytical models still need to be tested against
observations.

More recently,  the halo occupation model  has opened another  avenue to probe
the galaxy-dark matter connection (e.g.  Jing, Mo \& B\"orner 1998; Peacock \&
Smith 2000;  Berlind \&  Weinberg 2002; Cooray  \& Sheth 2002;  Scranton 2003;
Yang, Mo \& van den Bosch 2003;  van den Bosch, Yang \& Mo 2003; Yan, Madgwick
\&  White 2003;  Tinker \etal  2005; Zheng  \etal 2005;  Cooray 2006;  Vale \&
Ostriker 2006;  van den Bosch \etal  2007).  This technique  uses the observed
galaxy luminosity  function and  two-point correlation functions  to constrain
the average number  of galaxies of given properties that  occupy a dark matter
halo  of given  mass.  Although  this  method has  the advantage  that it  can
typically yield much  better fits to the data  than the semi-analytical models
or  numerical simulations,  one typically  needs to  assume a  somewhat ad-hoc
functional form to describe the halo occupation model.

A more direct  way of studying the galaxy-halo  connection is by using
galaxy groups,  provided that  these are defined  as sets  of galaxies
that reside  in the same  dark matter halo\footnote{In this  paper, we
  refer to a system of galaxies as a group regardless of its richness,
  including isolated galaxies (i.e.,  groups with a single member) and
  rich  clusters  of galaxies.}.   With  a  well-defined galaxy  group
catalogue,  one can  not  only  study the  properties  of galaxies  as
function  of  their  group   properties  (e.g.   Yang  \etal  2005c,d;
Collister  \& Lahav  2005; van  den Bosch  \etal 2005;  Robotham 2006;
Zandivarez \etal 2006; Weinmann \etal  2006a,b) but one can also probe
how dark matter haloes trace the large-scale structure of the universe
(e.g.  Yang \etal  2005b, 2006; Coil \etal 2006;  Berlind \etal 2007). 
During  the past  two  decades, numerous  group  catalogues have  been
constructed from various galaxy  redshift surveys, most noticeably the
CfA redshift  survey (e.g.  Geller  \& Huchra 1983), the  Las Campanas
Redshift Survey  (e.g.  Tucker \etal 2000), the  2-degree Field Galaxy
Redshift Survey  (hereafter 2dFGRS; Merch\'an \&  Zandivarez 2002; Eke
\etal 2004,  Yang \etal 2005a;  Tago \etal 2006; Einasto  \etal 2007),
the high-redshift DEEP2 survey (Gerke  \etal 2005), and the Two Micron
All Sky Redshift Survey  (Crook \etal 2007).  Various group catalogues
have also been constructed from the redshift samples selected from the
on-going Sloan  Digital Sky Survey  (hereafter SDSS): Goto  (2005) and
Berlind  \etal (2006)  used  a friends-of-friends  (FOF) algorithm  to
identify  groups in  the SDSS  Data  Release 2  (DR2; Abazajian  \etal
2004), Miller \etal  (2005) used the C4 algorithm  to find clusters in
the SDSS DR2, Weinmann \etal  (2006a) used the halo-based group finder
of Yang  \etal (2005a) to identify  groups in the  New York University
Value-Added Galaxy Catalogue (NYU-VAGC)  of Blanton \etal (2005) which
is also based on the SDSS DR2, and Merch\'an \& Zandivarez (2005) used
a FOF  algorithm to identify groups  in the SDSS  DR3 (Abazajian \etal
2005).   Group catalogues  have also  been constructed  from  the SDSS
photometric  data.   Goto  \etal  (2002) developed  a  cut-and-enhance
method and applied  it to the early SDSS  commissioning data.  Bahcall
\etal (2003) compared the properties of groups selected from the early
SDSS commissioning data with two different selection methods, a hybrid
matched filter method (Kim 2002)  and a ``maxBCG'' method developed by
Annis \etal (1999).  Lee (2004)  identified compact groups in the SDSS
Early  Data  Release  (EDR;  Stoughton \etal  2002).   More  recently,
Koester \etal  (2007) used the  ``maxBCG'' method to assemble  a large
photometrically selected  galaxy group catalogue from the  SDSS with a
sky-coverage of $\sim 7500 {\rm deg}^{2}$. Photometric catalogues also 
exist outside the SDSS (e.g. Gonzalez et al. 2001; Gladders \& Yee 
2005).

In a  recent paper,  Yang \etal (2005a)  developed a  halo-based group
finder that is optimized for grouping galaxies that reside in the same
dark matter halo.  Using mock galaxy redshift surveys constructed from
the  conditional luminosity function  model (see  Yang et  al.  2004),
they found  that this group  finder is very successful  in associating
galaxies according to their common dark matter haloes.  In particular,
the group  finder performs also  reliably for poor  systems, including
isolated galaxies  in small mass  haloes.  This makes  this halo-based
group finder ideally suited to study the relation between galaxies and
dark matter haloes over a wide dynamic range in halo masses. Thus far,
the halo-based group finder has  been applied to both the 2dFGRS (Yang
\etal  2005a) and to  the SDSS  DR2 (Weinmann  \etal 2006a).   In this
paper,  we apply  a  slightly  modified and  improved  version to  the
NYU-VAGC based on  the SDSS DR4. As the first in  a series, this paper
focuses on the selection process and the basic properties of the group
catalogue.   More detailed analyses  of the  group properties  and the
implications for halo occupation  statistics and galaxy formation will
be presented in forthcoming papers.

This paper  is organized as  follows.  Section \ref{sec_data}  gives a
brief  description  of   the  SDSS  data  used  in   this  paper.   In
Section~\ref{sec_algorithm}  we describe  the halo-based  group finder
and  the methods  to assign  halo masses  to the  groups.   In Section
~\ref{sec_catalogue} we present the  group catalogue based on the SDSS
DR4, and  study some of  its basic properties.  Finally,  we summarize
our results in Section~\ref{sec_conclusion}.  Unless stated otherwise,
we adopt a $\Lambda$CDM  cosmology with parameters that are consistent
with the three-year data release  of the WMAP mission (hereafter WMAP3
cosmology):   $\Omega_{\rm   m}   =  0.238$, $\Omega_{\Lambda}=0.762$,
$\Omega_{\rm  b}=0.042$,   $n=0.951$, $h=H_0/(100  \kmsmpc)=0.73$  and
$\sigma_8=0.75$ (Spergel \etal 2007).

\section{Galaxy Samples}
\label{sec_data}

The data used in this paper is taken from the Sloan Digital Sky Survey
(SDSS; York  \etal 2000), a joint  five-passband $(u,g,r,i,z)$ imaging
and  medium-resolution  ($R \sim  1800$)  spectroscopic survey.   More
specifically we make use of the New York University Value-Added Galaxy
Catalogue (NYU-VAGC; see  Blanton \etal 2005), which is  based on SDSS
DR4 (Adelman-McCarthy  \etal 2006) but  includes a set  of significant
improvements  over the  original  pipelines.  From  this catalogue  we
select all  galaxies in the Main  Galaxy Sample with  redshifts in the
range $0.01 \leq z \leq  0.20$ and with a redshift completeness $\calC
> 0.7$ (about $4\%$ of the galaxies have $\calC\le 0.7$).  This leaves
a grand  total of $362356$ galaxies with  reliable $r$-band magnitudes
and  with measured redshifts  from the  SDSS.  We  will refer  to this
sample of galaxies as Sample I.
\begin{figure*}
\plotone{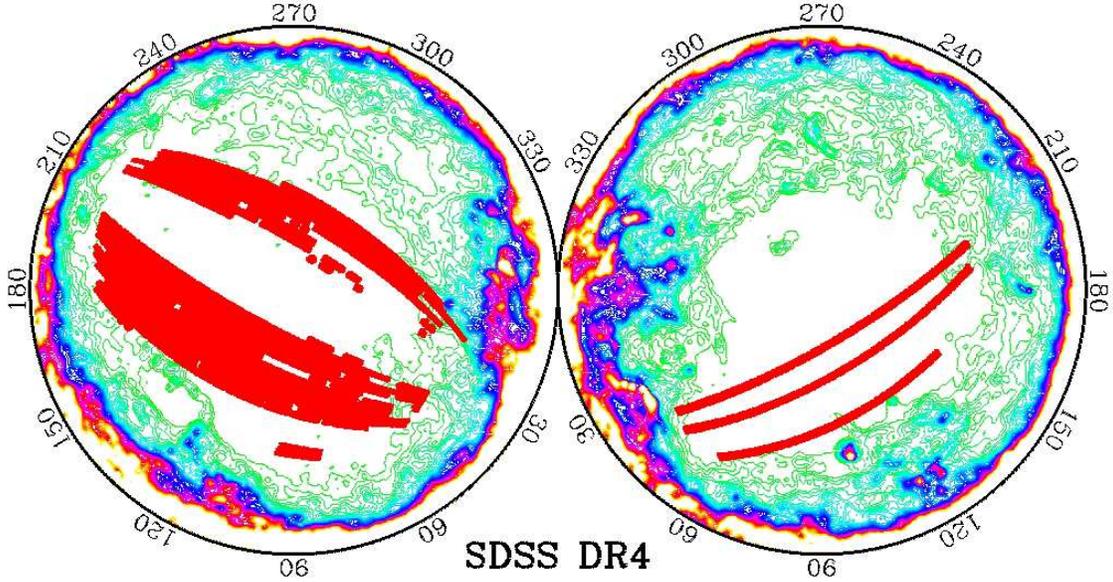}
\caption{The sky coverage of the SDSS DR4 galaxies in sample II, 
  overlaid on the galactic extinction contours of Schlegel, Finkbeiner
  \& Davis (1998).  Note that  the SDSS probes regions of low galactic
  extinction.}
\label{fig:sky_cov}
\end{figure*}

In addition, there are $7091$ galaxies with $0.01 \leq z \leq 0.20$ in
the NYU-VAGC  which have redshifts from alternative  sources: from the
2dFGRS (Colless  \etal 2001), from  the PSCz (Saunders \etal  2000) or
from the  RC3 (de Vaucouleurs \etal 1991)  \footnote{See Blanton \etal
  (2005) for  details.}.  Including these  galaxies results  in Sample
II,  with   a  total  of  $369447$  galaxies.    As  an  illustration,
Fig.~\ref{fig:sky_cov}  shows  the   sky  coverage  ($\sim  4514  {\rm
  deg}^2$)  of all  galaxies  in Sample  II  in Galactic  coordinates,
overlaid on  the galactic extinction contours  of Schlegel, Finkbeiner
\& Davis (1998).

The  two samples  described above  suffer from  incompleteness  due to
fiber collisions.  No two fibers on  the same SDSS plate can be closer
than 55 arcsec.  Although this fiber collision constraint is partially
alleviated by  the fact that neighboring plates  have overlap regions,
$\sim 7$ percent of all galaxies eligible for spectroscopy do not have
a  measured  redshift.   Hereafter  we  refer  to  these  galaxies  as
`fiber-collision' galaxies.  Since  fiber collisions are more frequent
in regions of high (projected)  density, they are more likely to occur
in richer groups,  thus causing a systematic bias that  may need to be
accounted for.   A simple  method of  doing so is  to assign  a galaxy
which lacks an observed redshift  due to fiber collisions the redshift
of  the galaxy  with  which it  collided.   As shown  in Zehavi  \etal
(2002),  roughly 60  percent of  the fiber-collision  galaxies  have a
redshift within $500  \kms$ of their nearest neighbor,  and for these
cases the  above procedure is  more than appropriate.   However, there
are also cases in which the fiber-collision galaxy has a true redshift
that is  very different  from that of  its nearest neighbor.   If the
fiber-collision galaxy is  assigned a redshift that is  too large, its
implied  luminosity will also  be too  large, and  can in  fact become
excessively large.   This in turn  can have dramatic  consequences for
our group finder.  To limit  the impact of these catastrophic failures
we remove the $\sim 1.0$  percent of all fiber-collision galaxies that
have  an  implied  absolute  magnitude  of  $\rmag  \leq  -22.5$  (see
eq.~[\ref{magn}] below).  In our  redshift interval, there are a total
of  $38672$ galaxies  with  an  assigned redshift  and  with $\rmag  >
-22.5$. Including these  galaxies results in Sample III,  with a total
of $408119$ galaxies.

In what  follows, we use  Sample II as  our main sample  for selecting
galaxy groups. For  completeness, we will also apply  our group finder
over samples I and III, and we will occasionally compare results based
on all three group catalogues. 

\subsection{Magnitudes and stellar masses}
\label{sec:magnmass}

For  each galaxy  we compute  the absolute  magnitude in  bandpass $Q$
using
\begin{equation}
\label{magn}
\Qmag = m_Q + \Delta m_Q - {\rm DM}(z) - K_Q - E_Q
\end{equation}
Here  ${\rm  DM}(z) =  5  \log\left[D_L/(\mpch)\right]  +  25$ is  the
bolometric  distance modulus calculated  from the  luminosity distance
$D_L$  using  a WMAP3  cosmology  with  $\Omega_{\rm  m} =  0.24$  and
$\Omega_{\Lambda}=0.76$.   $\Delta  m_Q$   is  the  latest  zero-point
correction  for  the  apparent  magnitudes, which  converts  the  SDSS
magnitudes to  the AB  system, and  for which we  adopt $\Delta  m_Q =
(-0.036,+0.012,+0.010,+0.028,+0.040)$   for  $Q=(u,g,r,i,z)$  (Michael
Blanton,  private communication).  All  absolute magnitudes  are $K+E$
corrected  to $z=0.1$.   For the  $K$  corrections we  use the  latest
version of `Kcorrect' (v4) described in Blanton \etal (2003a; see also
Blanton \&  Roweis 2007),  which we apply  to {\it all}  galaxies that
have meaningful  magnitudes and meaningful  redshifts, including those
that have redshifts from alternative  sources and those that have been
assigned  the  redshift  of  their  nearest  neighbor.   Finally,  the
evolution   corrections  to   $z=0.1$  are   computed  using   $E_Q  =
A_Q(z-0.1)$,   with   $A_Q   =  (-4.22,-2.04,-1.62,-1.61,-0.76)$   for
$Q=(u,g,r,i,z)$ (see Blanton \etal  2003a).  Note that these evolution
corrections imply that  galaxies were brighter in the  past (at higher
redshifts).

In  addition to  the absolute  magnitudes,  we also  compute for  each
galaxy  its stellar mass,  $M_*$. Using  the relation  between stellar
mass-to-light ratio and color of Bell \etal (2003), we obtain
\begin{eqnarray}
\label{eq:stellar}
\log\left[{M_* \over h^{-2}\Msun}\right] & = & -0.306 + 1.097
\left[^{0.0}(g-r)\right] - 0.1 \nonumber\\
& & - 0.4(\rrmag-4.64)\,,
\end{eqnarray}
Here  $^{0.0}(g-r)$ and $\rrmag$  are the  $(g-r)$ color  and $r$-band
magnitude $K+E$ corrected to $z=0.0$, $4.64$ is the $r$-band magnitude
of the Sun in the AB  system (Blanton \& Roweis 2007), and the $-0.10$
term  effectively implies  that we  adopt a  Kroupa (2001)  IMF (Borch
\etal 2006).   

For a  small fraction  of all galaxies,  the $g-r$ color  that results
from  the  photometric SDSS  pipeline  is  unreliable. These  galaxies
typically  have $g-r$ colors  that are  clearly unrealistic  (they are
catastrophic outliers  in the color-magnitude  distribution).  If this
is  not  accounted   for,  equation~(\ref{eq:stellar})  assigns  these
galaxies stellar  masses that are  unrealistically high or  low, which
can have a  dramatic impact on our group  finder (which assigns masses
to  the  groups  based  on  their  characteristic  stellar  mass;  see
Section~\ref{sec:groupmass} below). To  take account of these outliers
we  proceed  as  follows.   As   shown  by  Baldry  \etal  (2004)  the
distribution of  $(g-r)$ colors at  a given $r$-band magnitude  can be
well  approximated by  a  bi-Gaussian function,  representing the  red
sequence and the  blue cloud.  Following Li \etal  (2006) we therefore
fit bi-Gaussian  functions to the distribution of  $^{0.0}(g-r)$ for a
total of 118 bins in $\rrmag$.  As shown in Li \etal (2006) these fits
accurately capture the distribution of galaxies in the color-magnitude
plane. For  any galaxy that  falls outside the 3-$\sigma$  ranges from
the mean  color-magnitude relations of  both the red sequence  and the
blue cloud ($\sim 2\%$ of all  galaxies in Sample III), we compute its
stellar mass using the mean color of the red sequence (when the galaxy
is too red) or the blue  cloud (when the galaxy is too blue). Detailed
tests  have shown that  this prevents  any problems  with catastrophic
outliers.

\section{The Construction of the Group Catalogue}
\label{sec_algorithm}

\subsection{The Group Finder}
\label{sec:steps}

The group  finder adopted  here is similar  to that developed  in Yang
\etal (2005a).  The strength  of this group finder, hereafter referred
to as the  halo-based group finder, is that it  is iterative and based
on an  adaptive filter  modeled after the  general properties  of dark
matter haloes.   In addition, unlike the traditional  FOF method, this
group finder can also identify groups with only a single member, which
allows us to sample a wider dynamic range in group masses.
Note that various masses are used in our group finder and in the 
presentation. In order to avoid confusion, we list in 
Table~\ref{tab:name} the various masses that are used along 
with their definitions.

The halo-based group finder consists of the following main steps:

{\bf Step 1:  Find potential group centers.}  We  use a combination of
two  different  methods  to  identify  the centers  (and  members)  of
potential groups in redshift space.   First we use the traditional FOF
algorithm (e.g. Davis \etal 1985) with very small linking  lengths in 
redshift space to assign
galaxies into tentative groups that may represent the central parts of
groups.  The  linking lengths adopted are $\ell_z=0.3$  along the line
of sight, and $\ell_p=0.05$ in the transverse direction, both in units
of the mean  separation of galaxies at the  redshift in question.  The
geometrical, luminosity-weighted  centers of  all the FOF  groups thus
identified with two  members or more are considered  as the centers of
potential groups.  Next, for all  galaxies not yet linked to these FOF
groups, we treat them also as tentative centers of  potential groups.

{\bf Step 2: Determine the characteristic luminosity of each tentative
  group.}   In order  to  be able  to  meaningfully compare  different
groups,  we  define   the  group's  {\it  characteristic}  luminosity,
$L_{19.5}$, defined  as the combined  luminosity of all  group members
with $\rmag \leq -19.5$ (here again, all absolute magnitudes are $K+E$
corrected  to $z=0.1$).   For groups  with  redshifts $z  < 0.09$  all
galaxies  with $\rmag  \leq -19.5$  make the  flux limit  of  the SDSS
spectroscopic sample, and $L_{19.5}$ can be computed directly using
\begin{equation}
\label{Lchardirect}
L_{19.5} = \sum_i \frac{L_i}{\calC_i}\,,
\end{equation}
where  $L_i$ is  the luminosity  of  the $i^{\rm  th}$ member  galaxy,
$\calC_i$ is  the completeness of the  survey at the  position of that
galaxy, and the  summation is over all group  members with $\rmag \leq
-19.5$. For groups with $z>0.09$,  however, we need to correct for the
missing members with $\rmaglim \leq \rmag \leq -19.5$, with $\rmaglim$
the absolute magnitude limit at the redshift of the group. In this
case, we define the characteristic luminosity as
\begin{equation}
\label{eq:Ls_grp}
L_{19.5} =  {1\over f(L_{19.5},L_{\rm lim})}
\sum_i \frac{L_i}{\calC_i}\,,
\end{equation}
with $f(L_{19.5},L_{\rm  lim})$ a correction  factor ($0 < f  \leq 1$)
that  accounts  for  the  galaxies  missed  because  of  the  apparent
magnitude limit of the  spectroscopic survey.  The method of computing
$f(L_{19.5},L_{\rm lim})$ is described in \S~\ref{sec:compl} below.
  
{\bf Step  3: Estimate the mass,  size and velocity  dispersion of the
  dark matter halo associated  with each tentative groups.}  Using the
value of $L_{19.5}$  determined above and an assumption  for the group
mass-to-light  ratio, $M_h/L_{19.5}$, we  assign each  tentative group
with a halo  mass which we use in the following  steps to assign group
memberships. 

In the first iteration we simply adopt a constant mass-to-light ratio,
$M_h/L_{19.5}=500 h  \Msun/\Lsun$ for all groups.   For all subsequent
iterations, however,  we use the $M_h/L_{19.5}$  - $L_{19.5}$ relation
obtained from  the previous iteration  (using the method  described in
\S\ref{sec:groupmass}).  Because of this iterative technique the final
group catalogue is very  insensitive to the (fairly arbitrary) initial
guess of  $M_h/L_{19.5}=500 h \Msun/\Lsun$  (see Yang \etal  2005a for
detailed tests). Note that the halo masses in this step are estimated 
using the mass-to-light ratio, and agree well with the final masses 
to be estimated in Section~\ref{sec:groupmass}. 

Throughout  this paper  we  define  dark matter  haloes  as having  an
overdensity of  $180$. This implies,  for the WMAP3  cosmology adopted
here, a halo radius of
\begin{equation}\label{r180}
r_{180} = 1.26 \mpch \left( {M_h \over 10^{14} \msunh}\right)^{1/3} 
\, (1 + z_{\rm group})^{-1}
\end{equation}
where $z_{\rm group}$ is the redshift of the group center, and a 
line-of-sight velocity dispersion of
\begin{equation}
\label{veldispfit}
\sigma = 397.9 \kms \left( {M_h \over 10^{14} 
\msunh}\right)^{0.3214}\,.
\end{equation}
The latter  is a  fitting function that  accurately captures  the halo
mass dependence of the one-dimensional velocity dispersion as given by
equation~(14)  in   van  den  Bosch  \etal  (2004),   using  the  halo
concentrations of Maccio \etal (2007). 

\begin{deluxetable*}{ll}
\tabletypesize{\scriptsize}
\tablecaption{Various masses and their definitions \label{tab:name}}
\tablewidth{0pt}
\tablehead{
\colhead{Name} &
\colhead{Definition} 
}

\startdata
$M_*$ & stellar mass of a galaxy   \\
$M_{\rm stellar}$ & total stellar mass of group members with $\rmag\le -19.5$  \\
$M_h$ & true halo mass (unless stated otherwise) \\
$M_L$ & halo mass estimated using the ranking of $L_{19.5}$\\
$M_S$ & halo mass estimated using the ranking of $M_{\rm stellar}$
\enddata

\end{deluxetable*}
\begin{figure*}
\plotone{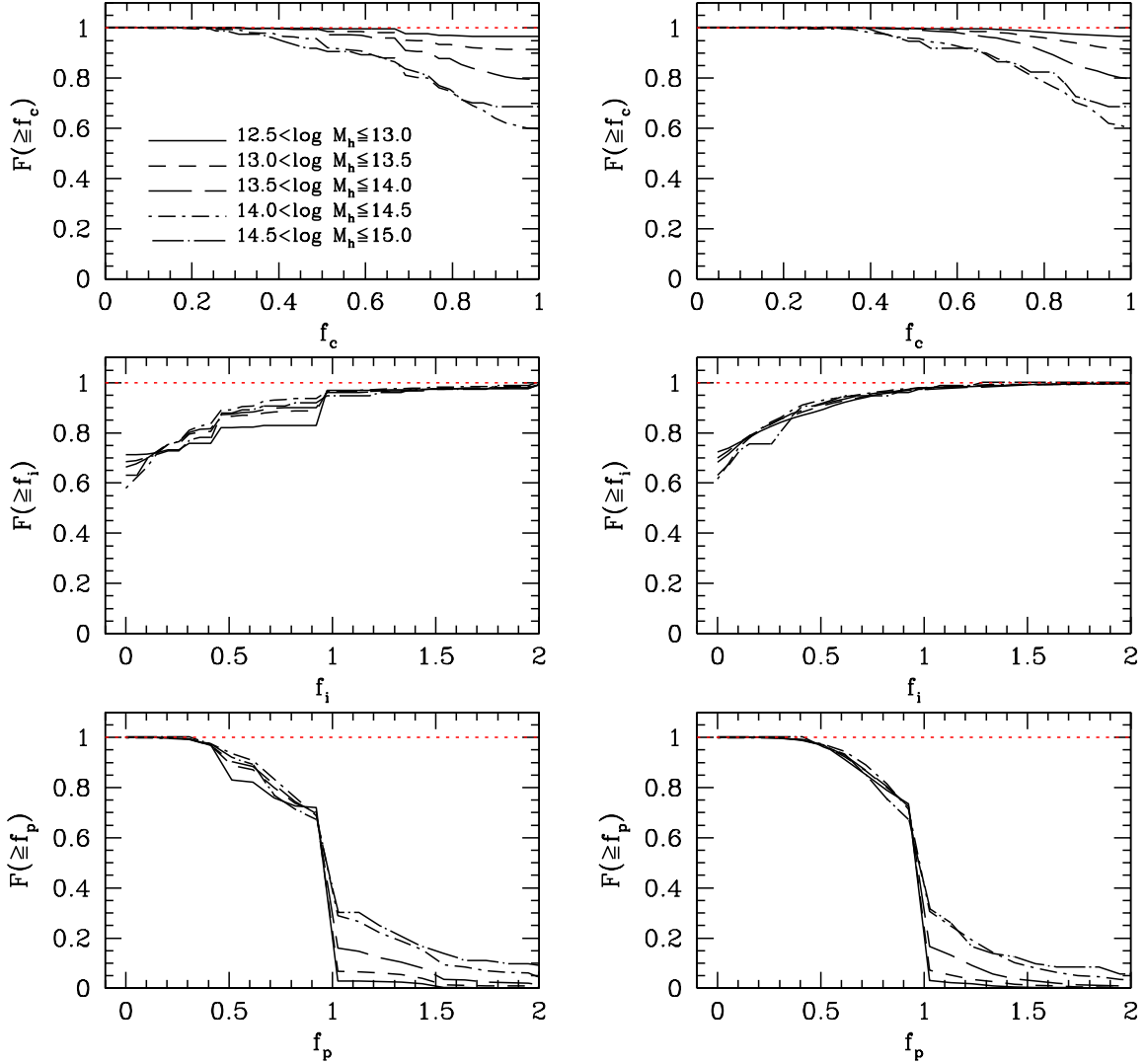}
\caption{The  upper, middle and lower panels show the cumulative 
  distributions  of completeness,  $f_{\rm c}$  (the fraction  of true
  members), contamination, $f_{\rm i}$, (the fraction of interlopers),
  and purity,  $f_{\rm p}$,  (ratio between the  true members  and the
  group members). See  text for the detailed definitions  of all three
  parameters.   In the left-  and right-hand  panels these  values are
  number  weighted and  luminosity weighted,  respectively.  Different
  lines show the  result for groups in haloes  of different masses, as
  indicated.  Results are plotted for  groups with at least 2 members,
  since groups with only 1 member have, by definition, $f_{\rm i}=0$.}
\label{fig:compl}
\end{figure*}

{\bf Step  4: Update group memberships using  halo information.}  Once
we have a group center and a tentative estimate of the size, mass, and
velocity  dispersion of  the halo  associated with  it, we  can assign
galaxies to this group using these halo properties.  If we assume that
the distribution of  galaxies in phase-space follows that  of the dark
matter  particles, the  number  density contrast  of  galaxies in  the
redshift space around  the group center (assumed to  coincide with the
center of the halo) at redshift $z_{\rm group}$ can be written as
\begin{equation}
P_M(R,\Delta z) = {H_0\over c} {\Sigma(R)\over {\bar \rho}} p(\Delta z) \,,
\end{equation}
where  $c$ is the  speed of  light, $\Delta  z =  z -  z_{\rm group}$,
$\bar{\rho}$ is  the average density  of Universe, and  $\Sigma(R)$ is
the projected surface density of  a (spherical) NFW (Navarro, Frenk \&
White 1997) halo:
\begin{equation}
\Sigma(R)= 2~r_s~\bar{\delta}~\bar{\rho}~{f(R/r_s)}\,,
\end{equation}
with $r_s$ the scale radius, 
\begin{equation}
\label{fx}
f(x) = \left\{
\begin{array}{lll}
\frac{1}{x^{2}-1}\left(1-\frac{{\ln
{\frac{1+\sqrt{1-x^2}}{x}}}}{\sqrt{1-x^{2}}}\right)   &  \mbox{if   $x<1$}  \\
\frac{1}{3}   &   \mbox{if   $x=1$}   \\   
\frac{1}{x^{2}-1}\left(1-\frac{{\rm
      atan}\sqrt{x^2-1}}{\sqrt{x^{2}-1}}\right) & \mbox{if $x>1$}
\end{array} \right.\,,
\end{equation}
and 
\begin{equation}
\bar{\delta} = {180 \over 3} {c_{180}^3 \over {\rm ln}(1 + c_{180}) -
c_{180}/(1+c_{180})}
\end{equation}
with $c_{180}=r_{180}/r_s$. The  function $p(\Delta z){\rm d}\Delta z$
describes the  redshift distribution of galaxies within  the halo, and
is assumed to have a Gaussian form,
\begin{equation}
p(\Delta z)=  {1 \over  \sqrt{2\pi}} {c \over  \sigma (1+z_{\rm  group})} \exp
\left [ \frac {-(c\Delta z)^2} {2\sigma^2(1+z_{\rm group})^2}\right ] \,,
\end{equation}
where   $\sigma$   is    the   rest-frame   velocity   dispersion   of
equation~(\ref{veldispfit}).  Thus  defined, $P_M(R,\Delta z)$  is the
three-dimensional  density contrast  in redshift  space.  In  order to
decide whether  a galaxy should be  assigned to a  particular group we
proceed  as follows.  For  each galaxy  we loop  over all  groups, and
compute the distance  $(R,\Delta z)$ between the galaxy  and the group
center, where  $R$ is  the projected distance  at the redshift  of the
group.   If $P_M(R,\Delta  z)  \ge B$,  with  $B=10$ an  appropriately
chosen background level (see Yang \etal 2005a), the galaxy is assigned
to the  group.  If  a galaxy can  be assigned  to more than  one group
according to this criterion, it is  only assigned to the one for which
$P_M(R,\Delta  z)$ is  the largest.   Finally, if  all members  in two
groups can be assigned to  one group according to the above criterion,
the two groups are merged into a single group.
\begin{figure}
\plotone{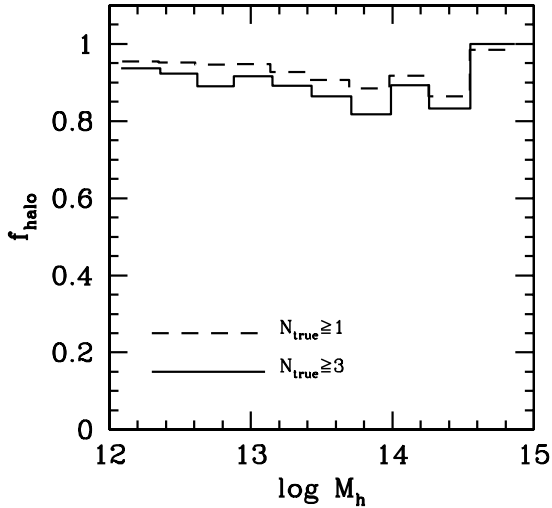}
\caption{The global completeness, $f_{\rm halo}$, defined as the
  fraction of haloes  in the MGRS whose brightest  member has actually
  been identified as  the brightest (central) galaxy of  its group, as
  function  of the true  halo mass  $M_h$. Results  are shown  for all
  haloes (dashed histogram)  and for those haloes with  at least three
  members in the MGRS (solid histogram).}
\label{fig:comp_halo}
\end{figure}

{\bf Step 5: Iterate.}  Using  the new memberships obtained in Step 4,
we re-compute the group centers and go back to Step 2.  This iterating
process  goes  on  until there  is  no  further  change in  the  group
memberships.   Next we use  the resulting  group catalogue  to compute
$f(L_{19.5},L_{\rm lim})$ and  the relation between $M_h/L_{19.5}$ and
$L_{19.5}$ and we go back to Step 1. We stop this iteration cycle once
the  $M_h/L_{19.5}$   -  $L_{19.5}$  relation   has  converged,  which
typically takes only 3 to 4 iterations.

\subsection{Completeness, Contamination and Purity of the Group
  Catalogues}
\label{sec:performance}

To test  its performance, we run  our halo-based group finder  over a detailed
mock galaxy  redshift survey  (MGRS) that  mimics the SDSS  DR4.  The  MGRS is
constructed  by populating  dark  matter haloes  in  numerical simulations  of
cosmological  volumes  with  galaxies  of different  luminosities,  using  the
conditional luminosity function  (CLF) model of van den  Bosch \etal (2007, in
preparation).   This CLF  describes  the halo  occupation  statistics of  SDSS
galaxies,  and  accurately  matches  the  SDSS  luminosity  function  and  the
clustering properties  of SDSS  galaxies as function  of their  luminosity. We
used a stack  of simulations with different resolutions  ($100$ and $300\mpch$
cubes with  $512^3$ dark  matter particles  each) to make  sure that  the mock
catalogue is complete  down to the SDSS magnitude limit  (see Yang \etal. 2004
for the stacking).   Next a MGRS is constructed mimicking  the sky coverage of
the SDSS  DR4 and  taking detailed  account of the  angular variations  in the
magnitude limits and completeness of the data (see Li \etal 2007 for details).
Methods  like this  are  becoming widespread  for  both understanding  cluster
detection (Yang et al., 2005a; Gerke  et al., 2005; Koester et al., 2007, Cohn
et al., 2007) and in quantifying selection functions (Rozo et al., 2007).

To assess  the performance  of the group  finder we follow  Yang \etal
(2005a) and  proceed as follows. For  each group, $k$, we  look up the
halo ID, $h_k$,  of the brightest group member,  and we define $N_{\rm
  t}$ as the total number of true members in the MGRS (with $0.01 \leq
z \leq 0.20$) that belong to  halo $h_k$, $N_s$ as the number of these
true members that  are selected as members of group  $k$, $N_i$ as the
number of interlopers (group members that belong to a different halo),
and $N_{\rm g}$ as the  total number of selected group members.  These
allow us to define, for each group, the following three quantities:
\begin{itemize}
\item The  completeness $f_{\rm c} \equiv N_{\rm s}/N_{\rm t}$
\item The  contamination $f_{\rm i} \equiv N_{\rm i}/N_{\rm t}$
\item The  purity $f_{\rm p} \equiv N_{\rm t}/N_{\rm g}$
\end{itemize}
Since $N_{\rm g} =  N_{\rm i} + N_{\rm s}$, we have  that $f_{\rm p} =
1/(f_{\rm c} + f_{\rm i})$. A purity $f_{\rm p} < 1$ implies that the
number of interlopers is larger than the number of missed true members, 
while $f_{\rm p}>1$ implies that the group is not complete ($f_{\rm  c} 
< 1$) and the number of missed true members is larger than the number 
of interlopers. Note that the identity
of the halo that belongs to a  group is solely based on the halo ID of
the  brightest group member.  Consequently, the  contamination $f_{\rm
  i}$ can be larger than unity.  An ideal, perfect group finder yields
groups with $f_{\rm c} = f_{\rm p} = 1$ and $f_{\rm i}=0$. In the case
of the halo-based group finder used here, the value for the background
level $B$ has  been tuned to maximize the average  value of $f_{\rm c}
(1 - f_{\rm i})$ (see Yang \etal 2005a).
\begin{figure*}
\plotone{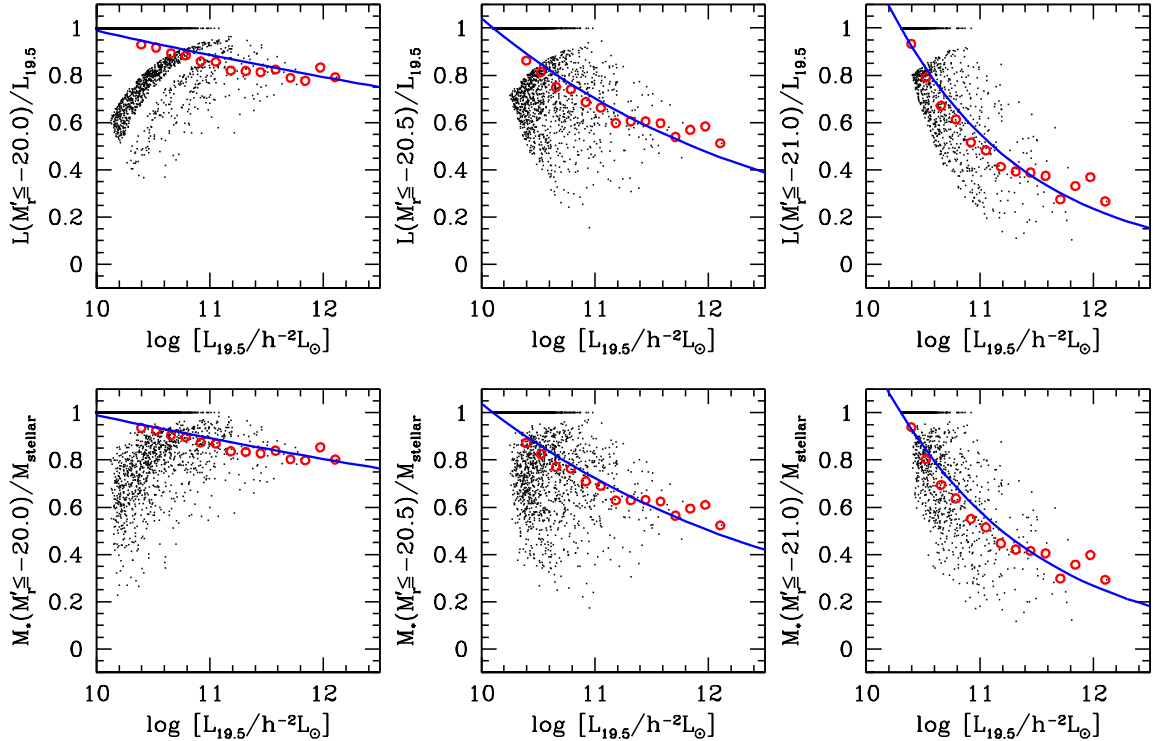}
\caption{{\it Upper panels:} The fraction of the characteristic
  luminosity $L_{19.5}$ that is  contributed by galaxies above a given
  magnitude limit as a  function of $L_{19.5}$.  Left-hand, middle and
  right-hand    panels    correspond    to   magnitude    limits    of
  $M_{r}^{\prime}\equiv   \rmag   =   -20.0,   -20.5$   and   $-21.0$,
  respectively.  Each  dot corresponds  to a group  in our  SDSS group
  catalogue  based  on Sample  II  with  $z<0.09$.   The open  circles
  indicate the mean fractions for a given bin in $L_{19.5}$, while the
  solid line  is the  exponential function that  best fits  these mean
  values,  and  which   defines  our  completeness  correction  factor
  $f(L_{19.5},  L_{\rm  min})$  discussed  in the  text.   {\it  Lower
    panels:} same  as the upper panels,  except that here  we plot the
  fraction   of  characteristic   stellar  mass,   $M_{\rm  stellar}$,
  contributed by galaxies above a given magnitude limit.}
\label{fig:f}
\end{figure*}

Results  obtained from  the  MGRS are  shown  in Fig.~\ref{fig:compl}.   Since
groups  with  a single  member  have  zero  contamination ($f_{\rm  i}=0$)  by
definition, results are  only shown for groups with a  richness $N\ge 2$.  The
upper   left-hand   panel  of   Fig.~\ref{fig:compl}   shows  the   cumulative
distributions  of   the  completeness  $f_{\rm   c}$.   Different  line-styles
correspond  to  groups of  different  true  halo  masses, as  indicated.   The
fraction of  groups with  100 percent completeness  (i.e., with  $f_{\rm c}=1$
depends on  group mass,  and ranges from  $\sim 95\%$  for low mass  groups to
$\sim 60\%$ for the most  massive clusters. Since
our group  finder is tuned to  maximize the average  value of $f_{\rm c}  (1 -
f_{\rm i})$, massive groups with larger velocity  dispersions have
larger $f_{\rm i}$ due to the contamination of foreground and background
galaxies. A compromise between $f_{\rm c}$ and $f_{\rm i}$ leads
to smaller $f_{\rm c}$ for more massive groups. 
Almost independent of group mass, we find that more  than $90\%$ of 
all groups have a  completeness $f_{\rm c} > 0.6$, while an average 
of $80\%$ of  all groups have $f_{\rm c} > 0.8$.  
The middle left-hand panel  of  Fig.~\ref{fig:compl}  shows  
the cumulative  distributions  of  the
contamination $f_{\rm i}$.  On average,  around $65\%$ of the groups have zero
contamination ($f_{\rm  i}=0$), while $\sim  85\%$ of the groups  have $f_{\rm
i}\le  0.5$, again  virtually  independent of  group  mass. These  interlopers
(contamination) are  either nearby  field galaxies or  the member  galaxies of
nearby massive groups, especially those along the line of sight. Finally,
the lower  left-hand panel  shows the cumulative  distributions of  the purity
$f_{\rm p}$, indicating that there are  on average as many groups with $f_{\rm
p} < 1$ as with $f_{\rm p}>1$. The break at $f_{\rm p}~1$ 
means that the number of recovered group members is about the same as 
the number of true members. Thus, the sharper the break is, 
the better. An ideal situation is a step function at $f_{\rm p}=1$.
In addition,  only a
negligibly small fraction  of groups have $f_{\rm p}<0.5$,  while only for the
most massive haloes is there a significant fraction ($\sim 10\%$) with $f_{\rm
p}>1.5$.

We also determine the  completeness, contamination and purity in terms
of the  total luminosity  rather than the  number of member  galaxies. 
The  corresponding  results are  shown  in  the  right-hand panels  of
Fig.~\ref{fig:compl}, respectively.   As one can see,  the results are
very similar  to those in terms  of the number of  members.  

As  a  final,  quantitative  assessment  of  the  performance  of  our
halo-based  group finder,  we examine  the {\it  global completeness},
$f_{\rm halo}$,  defined as the fraction  of haloes in  the MGRS whose
brightest  member  has  actually  been  identified  as  the  brightest
(central) galaxy of its group.  Fig.~\ref{fig:comp_halo} shows $f_{\rm
  halo}$ obtained from our CLF mock for haloes with $N_{\rm t} \geq 1$
(dashed lines)  and $N_{\rm t} \geq  3$ (solid lines)  as functions of
the true  halo mass.   As one can  see, the group  finder successfully
selects  more than  $90\%$ of  all the  true haloes  with  masses $\ga
10^{12}\msunh$ almost  independent of their  richness and with  only a
very weak dependence on halo mass.  Note that this does not imply that
$\sim 10\%$  of the  central galaxies in  dark matter haloes  have not
been selected  by the group  finder.  Especially for the  more massive
haloes, the vast majority of these central galaxies have been selected
as a group  member, but they are not the  brightest group member. This
can happen whenever  two nearby haloes are merged  into a single group
by the  group finder.  

\subsection{Completeness Corrections for the Characteristic Luminosity 
and Stellar Mass}
\label{sec:compl}

An  important   parameter  for   each  group  is   its  characteristic
luminosity,   defined   by   equation~(\ref{eq:Ls_grp}).   Since   the
correction   factor  $f(L_{19.5},   L_{\rm  min})$   depends   on  the
characteristic luminosity $L_{19.5}$ itself, it can only be determined
in an iterative  way.  In the first iteration of  our group finder, we
use
\begin{equation}
f(L_{19.5},L_{\rm lim}) = \frac{\int_{L_{\rm lim}}^{\infty} L\phi(L)dL}
{\int_{L_{\rm cut}}^{\infty} L\phi(L)dL}\,,
\end{equation}
where   $L_{\rm  cut}$   is   the  luminosity   that  corresponds   to
$\rmag=-19.5$, and  $\phi(L)$ is the galaxy  luminosity function, here
assumed to  be that  obtained by Blanton  \etal (2003b).   However, as
discussed  in Yang  \etal  (2005a), it  is  not reliable  to make  the
correction based on the assumption that the galaxy luminosity function
in groups  of a given  mass is  the same as  that of the  total galaxy
population.   After  all,   the  conditional  luminosity  function  of
galaxies in  groups varies significantly  with group mass  (Yang \etal
2005c; Zheng  \etal 2005).  Therefore, in the  following iterations we
use the  group catalogue of  the previous iteration  to self-calibrate
$f(L_{19.5},L_{\rm lim})$. To  do so, we first select  all groups with
$z  <  0.09$  (for  which  $f=1$)  and  compute  their  characteristic
luminosities.  Next, we use these  groups to determine the fraction of
the  characteristic luminosity  that is  contributed by  group members
with $L \geq  L_{\rm lim}$ for different values  of $L_{\rm lim}$. The
upper panels of Fig.~\ref{fig:f}  show the results for three different
values of  $L_{\rm lim}$ (corresponding  to $\rmaglim =  -20.0, -20.5$
and $-21.0$, from left to right). Next we determine the mean values of
these  fractions  as  function  of  $L_{19.5}$,  which  are  shown  in
Fig.~\ref{fig:f}  as open  circles, and  we  define $f(L_{19.5},L_{\rm
  lim})$ as the exponential function  that best fits these mean values
(shown as  solid lines in  Fig.~\ref{fig:f}). Note, however,  that the
scatter  around  these mean  values  is  fairly large.   Consequently,
despite the fact that our correction factors are self-calibrated in an
iterative way, they are only  accurate in a statistical sense, and are
not expected to be accurate for individual groups. As we will show in 
Section~\ref{sec:groupmass}, a considerable amount of scatter in the 
halo masses can be introduced by such correction. 
\begin{figure*}
\plotone{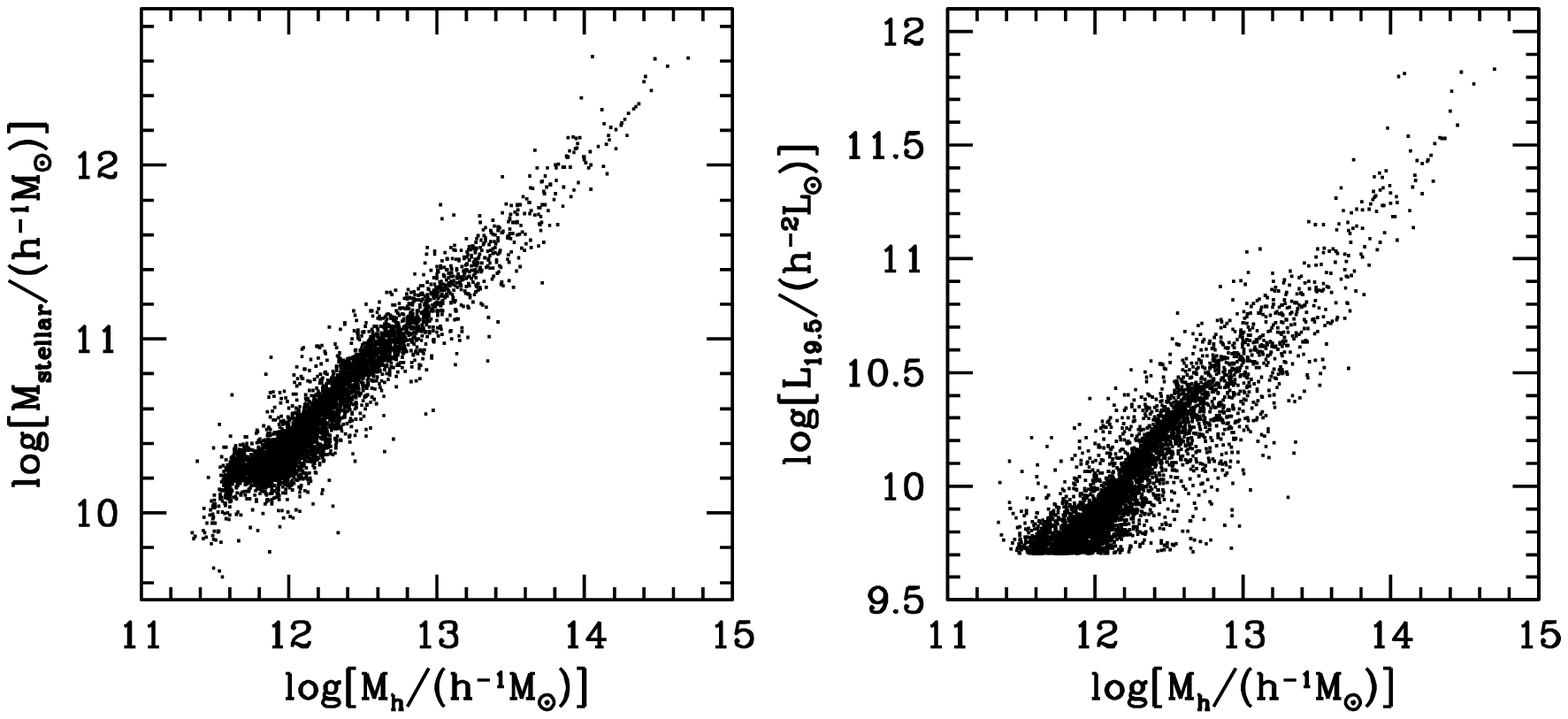}
\caption{The distributions  of  characteristic  stellar  mass  $M_{\rm
    stellar}$ (total stellar mass  of galaxies with $\rmag \le
  -19.5$ in a halo,  left-hand panel) and luminosity $L_{19.5}$ (total
  luminosity of  galaxies with  $\rmag \le  -19.5$ in  a halo,
  right-hand panel) as function of  halo mass, $M_h$.  The results are
  obtained  from  the semi-analytical  model  of  Kang  \etal (2005).  
  Obviously  both   $M_{\rm  stellar}$  and   $L_{19.5}$  are  tightly
  correlated with the halo mass. }
\label{fig:M*}
\end{figure*}

Similar to the characteristic luminosity defined above, we also define
a characteristic stellar mass
\begin{equation}
\label{eq:Ms_grp}
M_{\rm stellar} =  {1\over g(L_{19.5},L_{\rm lim})} 
\sum_i \frac{M_{*,i}}{\calC_i} \,,
\end{equation}
where as for  the characteristic luminosity the summation  is over all
group members  with $\rmag \leq -19.5$,  and $g(L_{19.5},L_{\rm lim})$
is  a  similar  correction  factor as  $f(L_{19.5},L_{\rm  lim})$  but
tailored  to the  stellar mass  rather than  the $r$-band  luminosity. 
Similar to $f$ these correction factors can be self-calibrated and the
results  are  shown in  the  lower-panels  of Fig.~\ref{fig:f}.   

\subsection{Correcting for Survey Edges}
\label{sec:edge}

An additional incompleteness effect that  needs to be accounted for is
due to the survey geometry. A group whose projected area straddles one
or more survey edges may have members that fall outside of the survey,
thus  causing  an  incompleteness,  which  in turn  affects  our  mass
estimate  of the  group.   The geometry  of  the survey  used here  is
defined as the region on  the sky where the SDSS redshift completeness
$\calC   >   0.7$,   and   is   indicated  by   the   red   areas   in
Fig.~\ref{fig:sky_cov}.  Clearly, this  geometry is fairly complicated
which can potentially have  a significant impact on various statistics
of the group  catalogue (cf.  Cooper \etal 2005;  Berlind \etal 2006). 
In order to correct for these edge effects we proceed as follows:

First, we estimate the mass  for each group using the method described
in Section~\ref{sec:groupmass} below  without taking edge effects into
account.   We  then  randomly   distribute  $200$  points  within  the
corresponding   halo  radius   $r_{180}$  (which   we   compute  using
Eq.~\ref{r180}). Next  we apply  the SDSS DR4  survey mask  and remove
those random  points that  fall outside of  the region where  $\calC >
0.7$.  For each group we  then compute the number of remaining points,
$N_{\rm remain}$, and we define $f_{\rm edge}=N_{\rm remain}/200$ as a
measure for the volume of  the group that lies within the survey
edges.   Finally we  multiply  $L_{19.5}$ and  $M_{\rm stellar}$  with
$1/f_{\rm edge}$ to  correct for the `missing members'  outside of the
edges of the survey. Tests with MGRSs show  that this correction works
well, except  for groups  with a small  $f_{\rm edge}$.   We therefore
discard those groups  with $f_{\rm edge} < 0.6$,  which removes (only)
$1.6\%$  of all  groups.  After  this correction  for edge  effect, we
re-calculate   the    mass   for   each   group    as   described   in
Section~\ref{sec:groupmass}.   The mass  difference  before and  after
this edge  effect correction is  relatively small: in most  cases less
than  $3\%$ and on  average less  than $10\%$.   Since this  change in
group  mass translates  only in  very  small changes  in $r_{180}$  no
iteration of this procedure is required.

\subsection{Estimating Group Masses}
\label{sec:groupmass}

An   important  aspect  of   each  galaxy   group  catalogue   is  the
determination of  the masses  of the groups.   Most studies  infer the
(dynamical) group  mass from the  velocity dispersion of  their member
galaxies.   However, the  vast majority  of the  groups in  our sample
contain only a  few members making a dynamical  mass estimate based on
its   members   extremely  unreliable.    Mass   estimates  based   on
gravitational lensing  (either strong or  weak) or on  X-ray emission,
also can  only be applied  to the most massive  systems.  Furthermore,
these latter  two methods require  high-quality data in excess  to the
information  directly  available  from  the redshift  survey  used  to
construct the group catalogue, rendering them impractical.

Rather,  we  estimate  the  group  masses  from  their  characteristic
luminosities or characteristic stellar masses.  This has the advantage
that (i) it is equally  applicable to groups spanning the entire range
in richness,  and (ii)  it does not  require any additional  data.  As
demonstrated in  Yang \etal  (2005a), the mass  of a dark  matter halo
associated  with  a  group   is  tightly  correlated  with  the  total
luminosity of  all member galaxies  down to some luminosity.   This is
further   illustrated  in   Fig.~\ref{fig:M*},  where   we   plot  the
correlations  between the  halo  mass, $M_h$,  and the  characteristic
stellar  mass $M_{\rm stellar}$  (left-hand panel)  and characteristic
luminosity, $L_{19.5}$ (right-hand panel) in the semi-analytical model
of Kang \etal (2005).   Clearly, both $M_{\rm stellar}$ and $L_{19.5}$
are   tightly   correlated   with   halo  mass,   with   the   $M_{\rm
  stellar}-M_{\rm  h}$  relation  being  slightly  tighter  than  that
between $L_{19.5}$  and $M_h$.  This is  expected, since $M_{\rm
  stellar}$ is less affected by  the current amount of star formation,
and suggests that the characteristic stellar mass is a somewhat better
mass indicator than the  characteristic luminosity. On the other hand,
the luminosities  are directly observed, while the  stellar masses are
derived quantities,  which creates additional  scatter.  Therefore, we
will compute  two mass estimates for  each group; $M_S$,  based on the
characteristic stellar mass $M_{\rm stellar}$, and $M_L$, based on the
characteristic luminosity $L_{19.5}$.  Throughout, we will compare all
results from the group catalogue for both mass estimates.
\begin{figure*}
\plotone{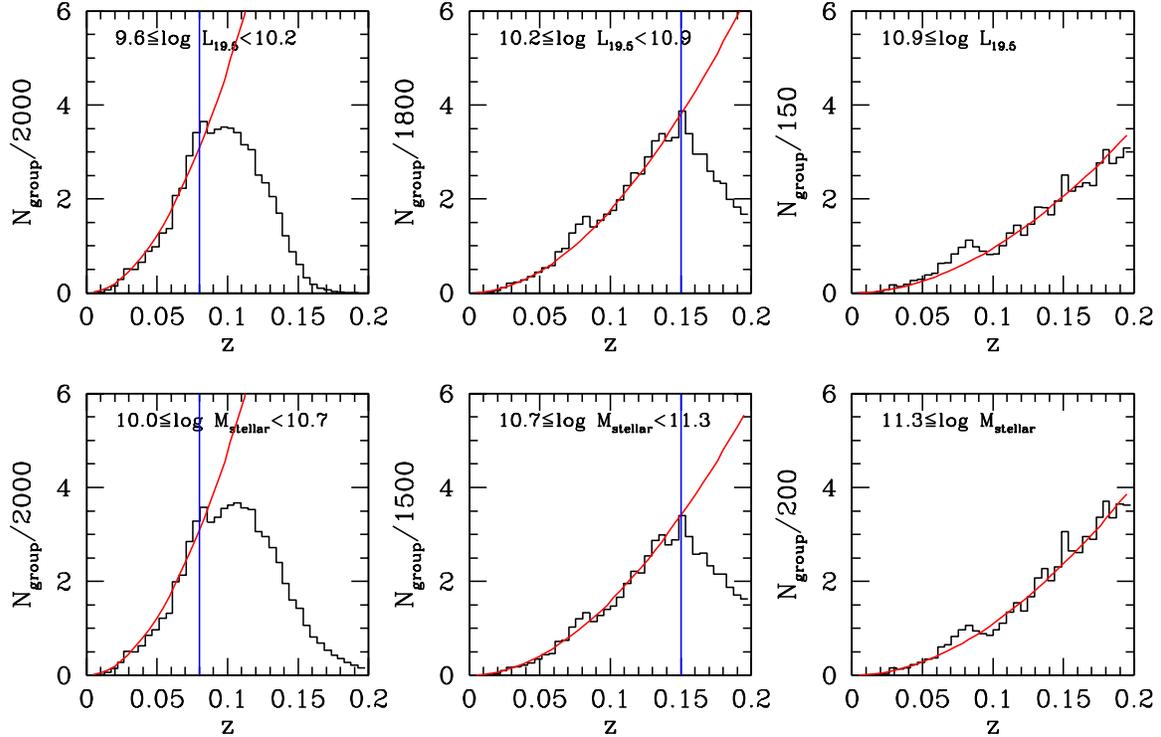}
\caption{The redshift distributions of groups for three bins in 
  characteristic    luminosity   $L_{19.5}$    (upper    panels)   and
  characteristic stellar  masses $M_{\rm stellar}$  (lower panels), as
  indicated. Solid  lines indicate the expected values  for a constant
  group number  density. Whenever the observed  distribution of groups
  starts  to systematically  drop  below this  line,  we consider  the
  sample incomplete. The vertical  lines indicate the redshifts out to
  which we  consider the samples  complete. In the  right-hand panels,
  the group samples are considered  complete out to the redshift limit
  of our galaxy sample ($z=0.2$).}
\label{fig:z_dist}
\end{figure*}

In order to convert the characteristic luminosities and stellar masses
to  halo masses, we  make the  assumption that  there is  a one-to-one
relation  between $L_{19.5}$  (or $M_{\rm  stellar}$) and
$M_h$. For  a given (comoving)  volume and a given halo mass function, 
$n(M_h)$, one can then link the characteristic luminosity
or stellar mass  to a halo mass by matching  their rank orders.  Note,
however, that this  only works for a group sample  that is complete in
$L_{19.5}$  or $M_{\rm stellar}$.   In Fig.~\ref{fig:z_dist},  we plot
the  redshift distributions  of groups  in three  different  ranges of
$L_{19.5}$  (upper  panels)  and  $M_{\rm stellar}$  (lower  panels).  
Comparing these distributions with that expected for a constant number
density (shown as  the solid line), we obtain  the rough redshifts out
to   which   these  different   samples   are   complete.   In   Table
~\ref{tab:ranking}, we list the  redshift limits thus obtained for the
three different  bins of  mass indicators, along  with the  numbers of
groups in each of the complete samples.  Only groups in these complete
samples  are used  in  the ranking;  the  masses of  other groups  are
estimated by  linear interpolation of the relations  between $M_h$ and
each  of the  mass  indicators  obtained from  the  complete samples.  
Because of the  particular volume limited samples used,  we can assign
group masses down to $10^{11.6} \msunh$.
\begin{deluxetable*}{lcccccc}
\tabletypesize{\scriptsize}
\tablecaption{Complete Samples used for Mass Ranking \label{tab:ranking}}
\tablewidth{0pt}
\tablehead{
\colhead{Redshift} &
\colhead{$\log L_{19.5}$} &
\colhead{Groups} &
\colhead{$\log M_{\rm stellar}$} &
\colhead{Groups} \\
 & & Samples I/II/III & & Samples I/II/III  \\
 (1) & (2) & (3) & (4) & (5) 
}

\startdata
$0.01\le z\le 0.20$ & $\ge 10.9$  & 7583/7683/8409  & $\ge 11.3$  & 11740/11851/12012   \\
$0.01\le z\le 0.15$ & [10.2~10.9] & 75120/75306/66001 & [10.7~11.3] & 53248/53377/47953  \\
$0.01\le z\le 0.08$ & [9.6~~10.2] & 33898/33939/36038 & [10.0~10.7] & 32739/32789/32702
\enddata

\tablecomments{Properties  of  the  three  complete  samples  used  to
  estimate group mass via  the ranking of characteristic luminosity or
  characteristic stellar mass. Column  (1) lists the redshift range of
  each sample. Columns  (2) and (4) lists the  corresponding ranges in
  $\log L_{19.5}$  and $\log M_{\rm  stellar}$, respectively. Finally,
  columns (3) and (5) lists the corresponding numbers of groups in the
  catalogues based on galaxy samples I, II and III, respectively. }
\end{deluxetable*}
\begin{figure*}
\plotone{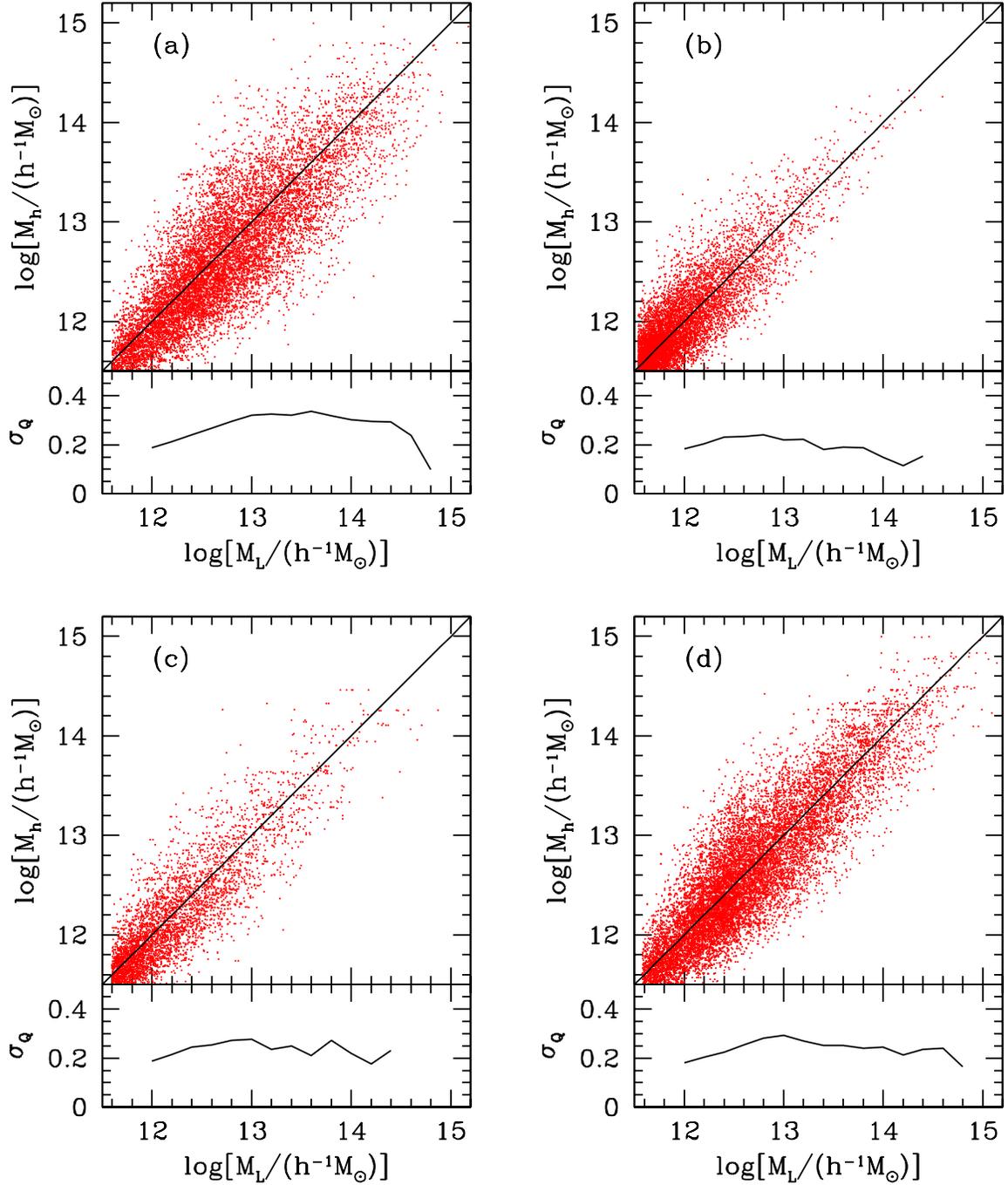}
\caption{{\it Panel (a):}  comparison between the   assigned halo mass
  $M_{L}$,  based on the  characteristic group  luminosity $L_{19.5}$,
  and the true halo mass  $M_h$. These results are obtained from
  the mock group catalogue constructed  from our MGRS. The small panel
  plots    the    standard     deviation    in    $Q$    defined    by
  equation~(\ref{Qparam}),  and reflects  the amount  of  scatter with
  respect to the  line of equality $M_L = M_h$ (shown  as a solid line
  in the scatter plot). {\it Panel (b):} Same as panel (a) except that
  here we show the comparison between $M_h$ and the assigned halo mass
  $M_L$  estimated  from  the  ranking of  the  characteristic  group
  luminosity $L_{19.5}$  obtained {\it  directly} from the  true group
  members  in the  simulation box  used to  construct the  MGRS.  {\it
    Panel  (c):} same as  panel (a)  but this  time only  plotting the
  results for groups  with $z\le 0.09$ for which one  does not need to
  correct  $L_{19.5}$ for missing  members. {\it  Panel (d):}  same as
  panel (a) but where we  have mimicked a perfect group finder without
  interlopers ($f_{\rm  i}=0$) and with a completeness  $f_{\rm c}=1$. 
  See text for  a detailed discussion. }
\label{fig:M_true}
\end{figure*}

Clearly,  the   assumption  of  a  one-to-one   relation  between  the
characteristic  luminosity  or  stellar  mass  and the  halo  mass  is
oversimplified.   In reality,  these relations  contain  some scatter,
which  results  in errors  in  our  inferred  group masses.   However,
detailed tests with mock galaxy  redshift surveys have shown that this
method  nevertheless  allows for  a  very  accurate  recovery of  {\it
  average}  halo  occupation  statistics.   In particular,  the  group
finder   yields   average   halo   occupation  numbers   and   average
mass-to-light ratios  that are in  excellent agreement with  the input
values  (Yang  \etal  2005b;  Weinmann \etal  2006a).   An  additional
shortcoming of our method is it requires the halo mass function, which
is  cosmology dependent (e.g. Sheth, Mo \& Tormen 2001; Warren et al.
2006) \footnote{Throughout we  compute the  halo mass
  function using the formulae given in Warren et al. (2006) with
  the  transfer  function given  by  Eisenstein  \&  Hu (1998),  which
  properly accounts for the effects of baryons.}.  However, as we will
show in  Section~\ref{sec:mtol}, it is  extremely easy to  convert the
group masses to  another cosmology, without having to  rerun the group
finder.

In order to further assess the reliability of the halo masses assigned
to individual  groups, we use  the mock group catalogue  obtained from
the CLF-based mock.  Following the procedure described above we assign
each  (mock) group  a halo  mass  $M_L$ based  on its  ranking of  the
characteristic  group luminosity  $L_{19.5}$.  The  top-left  panel of
Fig~\ref{fig:M_true}  shows the  $M_L$ thus  obtained versus  the true
halo mass,  $M_h$, defined as  the mass of the  dark matter halo
that hosts the brightest group galaxy.  In  order  to quantify  the
scatter  with  respect to  the  line of  equality  ($M_L  = M_h$),  we
determine for each group the quantity
\begin{equation}
\label{Qparam}
Q \equiv {1 \over \sqrt{2}} \left[\log(M_L) - \log(M_h) \right]
\end{equation}
and  measure the standard  deviation, $\sigma_Q$,  in several  bins of
$\left[\log(M_L) +  \log(M_h) \right]/2$.   The results, shown  in the
small panel, indicate  that the scatter is $\sim  0.35$ dex for groups
with $10^{13} \msunh \la M_L  \la 10^{14.5} \msunh$, dropping to $\sim
0.2$ dex at the high and low mass ends.

There are several factors that  contribute to this scatter.  The first
is the intrinsic scatter in the relation between the halo mass and the
true   value   of  $L_{19.5}$.    The   upper   right-hand  panel   of
Fig.~\ref{fig:M_true} shows  the relation  between the true  halo mass
and  the assigned mass  based on  the ranking  of the  true $L_{19.5}$
obtained  from the  CLF  mock before  incorporating any  observational
effects (e.g.   magnitude limit, incompleteness and  survey boundary). 
In other  words, we  measure $L_{19.5}$ using  all mock  galaxies with
$\rmag  \leq -19.5$, independent  of whether  those galaxies  would be
incorporated  in the  mock survey  or not.   The resulting  scatter is
about 0.2 dex,  similar to that of the  semi-analytical model shown in
the right-hand panel of Fig.~\ref{fig:M*}.

The  second   source  of  scatter  owes  to   the  incompleteness  and
contamination  introduced by  our group  finder.  The  lower left-hand
panel of Fig.~\ref{fig:M_true} shows the relation between the assigned
mass and  the true mass for groups  with $z < 0.09$.   As discussed in
Section~\ref{sec:compl}, the characteristic luminosity of these groups
does not need to be  corrected for incompleteness due to the magnitude
limit of the  survey (i.e., all galaxies with  $\rmag \leq -19.5$ make
the  magnitude  limit  of  the  survey).  The  scatter  here  is  only
marginally larger  than the intrinsic  scatter shown in  the top-right
panel,  suggesting that  the group  finding algorithm  by  itself only
introduces a very small amount of uncertainty in the assigned masses.
\begin{figure*}
\plotone{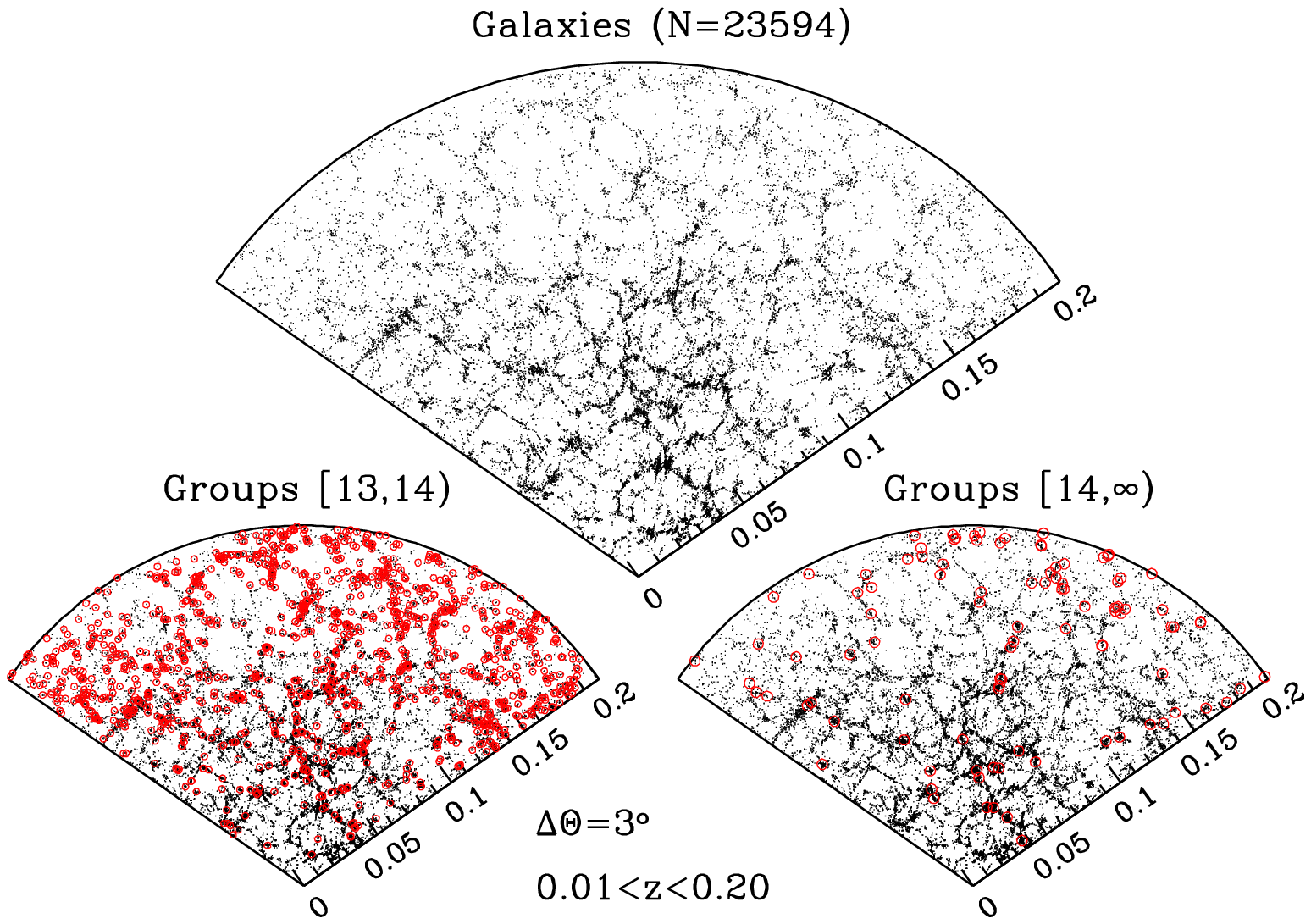}
\caption{The large wedge shows the distribution of a subset of
  SDSS DR4 galaxies in a $3^{\rm  o}$ slice in the south galactic pole
  region of the SDSS. These  distributions are repeated in the smaller
  wedges, where  we overplot, as  (red) open circles, the  groups with
  assigned  masses  in the  range  $10^{13}\msunh$ to  $10^{14}\msunh$
  (lower-left wedge)  and $>10^{14}\msunh$ (lower-right  wedge).  Note
  the halo masses  used in this plot are obtained  from the ranking of
  the characteristic luminosity $L_{19.5}$.}
\label{fig:slice}
\end{figure*}

The final source  of scatter in the assigned group  masses owes to the
fact  that  for  groups  with  $z  > 0.09$  we  need  to  correct  the
characteristic luminosity for  the group members that do  not meet the
magnitude limit of  the survey. As shown in  Fig.~\ref{fig:f} this can
introduce a  considerable amount of  scatter. To assess its  impact on
the inferred halo masses we  proceed as follows. We group all galaxies
in the mock SDSS DR4 according  to the halo to which they belong. This
resulting  `group  catalogue'  has,  by construction,  a  completeness
$f_{\rm  c}=1$, an  interloper fraction  $f_{\rm i}=0$,  and  a purity
$f_{\rm p}=1$.  For  each group in this perfect  catalogue we estimate
the characteristic  luminosity $L_{19.5}$: for groups with  $z < 0.09$
we  simply sum  the  luminosities  of all  galaxies  with $\rmag  \leq
-19.5$, while for  groups with $z>0.09$ we use  the correction factors
$f(L_{19.5},L_{\rm  lim})$ as  described  in Section~\ref{sec:compl}.  
Finally, we  assign each group  a mass $M_L$  based on the  ranking of
$L_{19.5}$  as described  above. The  lower right-hand  panel  of Fig~
\ref{fig:M_true}  plots the resulting  $M_L$ as  function of  the true
halo mass $M_h$. The scatter is $\sim 0.25$ dex, comparable to that in
the lower left-hand panel.

We therefore conclude that the majority of the scatter in the relation
between  the true  and  assigned  halo masses  owes  to the  intrinsic
scatter  in the  relation  between halo  mass  and its  characteristic
luminosity.   The fact  that the  group finder  is not  perfect (i.e.,
suffers  from interlopers  and  incompleteness) and  that  we need  to
correct the characteristic luminosity for members that do not make the
magnitude  limit   of  the  survey,  only  adds   a  relatively  small
contribution to the total scatter.
\begin{deluxetable*}{lcccccc}
\tabletypesize{\scriptsize}
\tablecaption{Number of Galaxies and Groups in the SDSS DR4 \label{tab:number}}
\tablewidth{0pt}
\tablehead{
\colhead{Samples} &
\colhead{Galaxies} &
\colhead{Groups} &
\colhead{$N = 1$} &
\colhead{$N = 2$} &
\colhead{$N = 3$} &
\colhead{$N > 3$} \\
(1) & (2) & (3) & (4) & (5) & (6) & (7) 
}

\startdata
Sample I   &  362356 &  295992 &  266763 &   19522 &    4511 &    5196\\
Sample II  &  369447 &  301237 &  271420 &   19868 &    4619 &    5330\\
Sample III &  408119 &  300049 &  250492 &   33537 &    7848 &    8172
\enddata

\tablecomments{For each of the three samples, columns (2) and (3) 
  list the number  of galaxies and of groups, respectively.   
  In addition,  columns (4)--(7)  list the  numbers of
  groups with $1$, $2$, $3$, and more than $3$ members.}

\end{deluxetable*}

\begin{figure*}
\plotone{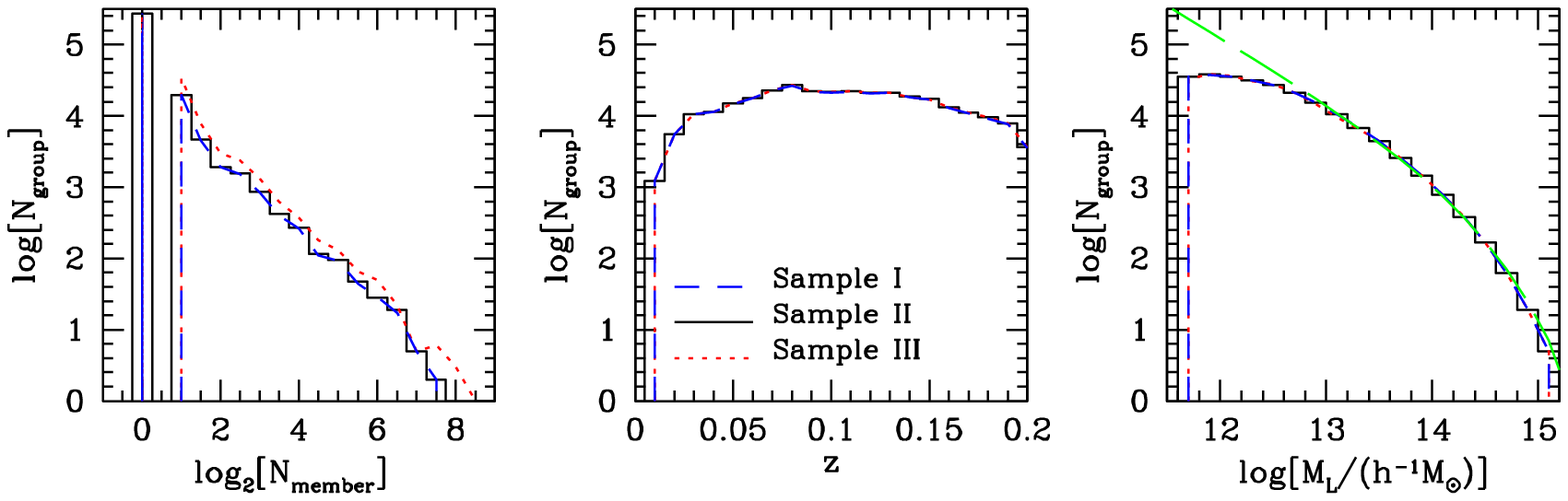}
\caption{The number of groups as function of the number of group
  members  (left-hand  panel),   group  redshift  (middle  panel)  and
  assigned halo mass (right-hand  panel). The dashed lines, histograms
  and dotted lines show the  results for the group catalogues based on
  Samples  I,  II  and  III,  respectively.  For  comparison,  in  the
  right-hand  panel we also  plot the  theoretical halo  mass function
  (long-dashed line). For $\log (M_L/\msunh)\ga 13$ the  mass function 
  of the groups is in excellent agreement with  this  theoretical mass
  function, indicating that our group sample is complete for this mass
  range. For  lower mass  groups, the sample  is only complete  out to
  lower redshifts (cf.  Table~1 and Fig.~\ref{fig:z_dist}).  Note that
  in the right-hand panel the group masses have been assigned based on
  their characteristic luminosity $L_{19.5}$. Using the characteristic
  stellar  mass,  $M_{\rm  stellar}$  instead  results  in  an  almost
  identical plot.}
\label{fig:N}
\end{figure*}

\section{Basic Properties of the Group Catalogue}
\label{sec_catalogue}

Application of  our halo-based group finder  to the SDSS  DR4 data set
described in Section~\ref{sec_data}  results in $295992$, $301237$ and
$300049$ groups for samples I, II and III, respectively. In
what  follows  we present  a  few  global  properties of  these  group
catalogues.

Table~2 lists,  for each  of the three  samples, the number  of groups
with 1,  2, 3, and more than  3 members: clearly, the  majority of the
groups contain only a single member.  Note also that sample III yields
many more systems with richness $N  \ge 2$ than samples I and II; this
is simply  due to the  fact that almost  all $38672$ galaxies  with an
assigned redshift  are members  of such systems.   As shown  in Zehavi
(2002), about 40\%  of these assigned redshifts have  an error of more
than $500\kms$.  This means that  in most cases these  galaxies should
not  have been  assigned  to the  group  in question  (i.e., they  are
interlopers),  which obviously  causes a  systematic bias  towards too
many members per  group. On the other hand, not  taking account of the
galaxies lost because of fiber  collisions results in an opposite bias
towards too few  members per group. We can assess  the impact of these
biases  on the  group  catalogue by  comparing  results obtained  from
Samples II  and III.  We  will come back  to this issue later  in this
section.

As  an illustration, Fig.~\ref{fig:slice}  shows the  distributions of
galaxies and groups in a $3^\circ$ slice.  As expected, massive groups
are located in  the denser regions of the  galaxy density field, while
groups  with  lower  masses   are  more  diffusely  distributed.   The
clustering properties of these  groups directly reflect the clustering
properties of  dark matter  haloes, and can  thus be used  to directly
probe the mass dependence of the halo bias (cf. Yang \etal 2005b; Coil
\etal 2006; Berlind \etal 2007).  We defer a more detailed analysis of
the  clustering  properties  of  the  groups in  the  SDSS  DR4  group
catalogues presented here to a forthcoming paper.
  
Fig.~\ref{fig:N} plots  the number  of groups as  a function  of group
richness (left  panel), redshift (middle panel), and  halo mass (right
panel). Group redshift is estimated using the luminosity-weighted 
average of all member galaxies. 
 Dashed,  solid and dotted histograms correspond  to the group
catalogues based on  Samples I, II and III,  respectively.  As already
mentioned above,  groups obtained  from Sample III  are systematically
richer than  those obtained from  the other two samples,  which simply
owes to the fact that all galaxies with an assigned redshift are group
members. However, as is evident from the middle and right-hand panels,
the redshift distributions and halo  mass functions of all three group
samples are extremely similar: although the inclusion of galaxies with
assigned  redshifts  changes  the   richness  of  the  systems,  their
redshifts   and  inferred  masses   are  virtually   unaffected.   The
long-dashed line  in the right-hand panel shows  the theoretical mass
function of dark matter haloes over the redshift range $0.01 \le z \le
0.20$.   The  mismatch  between  the  group  mass  function  and  this
theoretical halo mass function  at $\log [M_{L}/\msunh] \la 10^{12.8}$
is  caused by  the  incompleteness  of the  group  catalogue shown  in
Fig~\ref{fig:z_dist} and discussed in Section~\ref{sec:groupmass}.  If
we  would only  plot the  mass functions  for groups  in  the complete
samples of Table~1, they would, by construction, perfectly match their
theoretical equivalent.
\begin{figure*}
\plotone{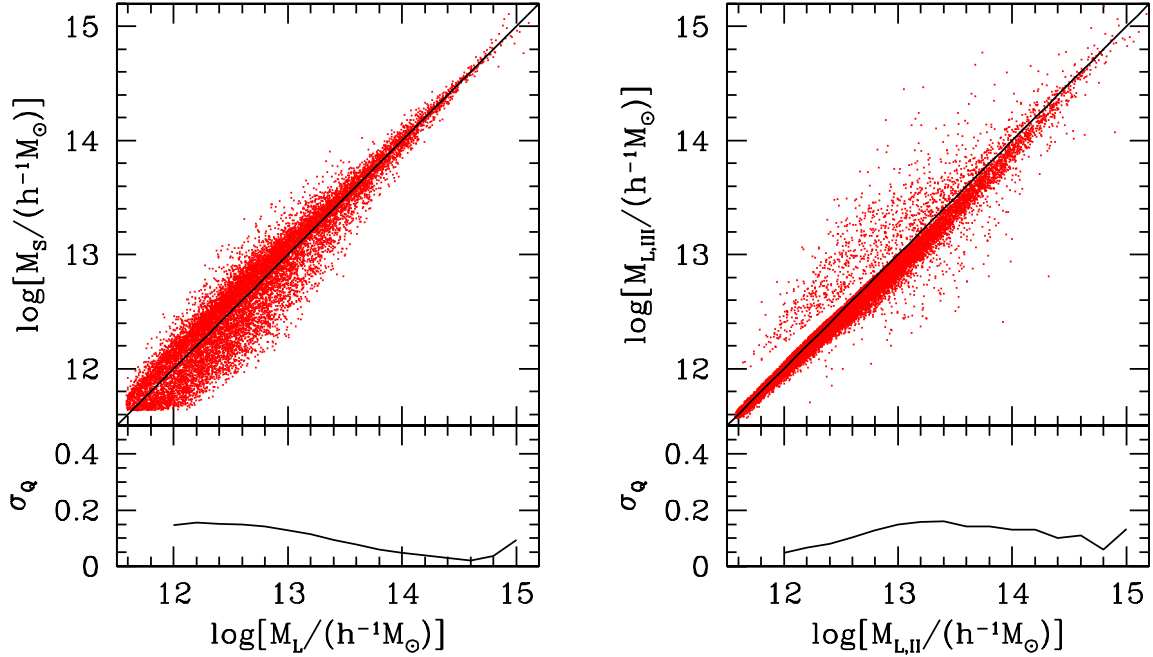}
\caption{{\it Left-hand panel:} Comparison of the   group  masses 
  $M_L$ and  $M_S$, obtained using the two  different mass indicators,
  $L_{19.5}$  and $M_{\rm stellar}$,  respectively. As  expected, both
  masses agree extremely well: the standard deviation in $Q$, shown in
  the small panel and  defined by equation~(\ref{Qparam}) is less than
  0.05 dex  at the massive end.   At the low mass  $\sigma_Q \sim 0.1$
  dex, due to the smaller  average number of galaxies per group. {\it
    Right-hand panel:} The halo mass  assigned to a group in Sample II
  versus the halo mass of the corresponding group in Sample III.  Here
  a group  in Sample III is defined  as the counterpart of  a group in
  Sample II if, and only if, it has the same brightest galaxy. Roughly
  $95\%$ of all groups in Sample II with $M_L\ga 10^{12}\msunh$ have a
  counterpart in Sample III, and for more than $90\%$ of these systems
  the difference  in the assigned group  mass is smaller  than $50\%$. 
  There  is  a small  number  of  outliers, but  they  do  not have  a
  significant  effect on  the  overall statistical  properties of  the
  group samples.}
\label{fig:IIandIII}
\end{figure*}

As discussed in Section~\ref{sec:groupmass}, group masses are obtained
down to  $10^{11.8} \msunh$ using  two different mass  indicators; the
characteristic  luminosity $L_{19.5}$  and the  characteristic stellar
mass     $M_{\rm     stellar}$.      The    left-hand     panel     of
Fig.~\ref{fig:IIandIII} compares  the inferred group  masses $M_L$ and
$M_S$, obtained using $L_{19.5}$  and $M_{\rm stellar}$, respectively. 
Overall, both  halo masses  agree very well  with each other,  with an
average scatter that decreases from $\sim 0.1$ dex at the low mass end
to $\sim 0.05$  dex at the massive end. This  scatter is expected, and
mainly  reflects that galaxies  of a  given luminosity  have different
colors, and therefore different  (inferred) stellar masses. The effect
is  somewhat  larger  for  lower  mass  groups  simply  because  their
characteristic mass  and luminosity are dominated by  a smaller number
of galaxies.

Finally, to assess  how the uncertainties in the  correction for fiber
collisions affect the group catalogue, we compare the masses of groups
in Sample  II with those  of its counterparts  in Sample III.   Here a
group in Sample III is defined as the counterpart of a group in Sample
II  if it  has the  same brightest  (central) galaxy.   We can  find a
Sample III  counterpart for $\sim  95\%$ of all  groups in Sample  II. 
There are  two main reasons  why a group  may not have a  counterpart. 
First  of  all, in  about  $3\%$  of the  groups  in  Sample III,  the
brightest group member is actually a galaxy with an assigned redshift,
so  that this  group can  not  have a  counterpart in  Sample II.   In
addition,  about $1\%$ of  the groups  in sample  II merge  with other
(nearby) groups  when the additional galaxies  with assigned redshifts
are used.   Consequently, in terms  of their brightest  galaxies, some
groups in Sample II have disappeared  in Sample III (i.e., do not have
a  counterpart),  while  others  have suddenly  increased  their  mass
substantially because  they have now  merged with another  group.  The
right-hand panel of Fig.~\ref{fig:IIandIII} plots the relation between
the  assigned halo  mass  of a  group in  Sample  II and  that of  its
counterpart in Sample  III. The relation is extremely  tight: for more
than $90\%$ of  the systems the difference in  the assigned group mass
is smaller than $50\%$, at  any given mass. This emphasizes once again
that although  the incompleteness due  to fiber collisions can  have a
significant effect on  the richness of individual groups,  it does not
have a significant  impact on which systems are  selected as groups or
on  their assigned  masses.  Although  the standard  deviation  in $Q$
(defined by equation~[\ref{Qparam}] but  with $M_L$ and $M_h$ replaced
by $M_{L,II}$ and $M_{L,III}$, respectively) reaches up to $\sim 0.25$
dex, this largely  owes to a very small fraction  of outliers that are
clearly  visible  in  the   scatter  plot.   In  particular,  one  can
distinguish  a   second  `sequence'  of  systems   with  $M_{L,III}  >
M_{L,II}$: this  corresponds to the $1\%$  of the groups  in Sample II
mentioned above that  have merged due to the  additional galaxies with
assigned redshifts.
\begin{figure*}
\plotone{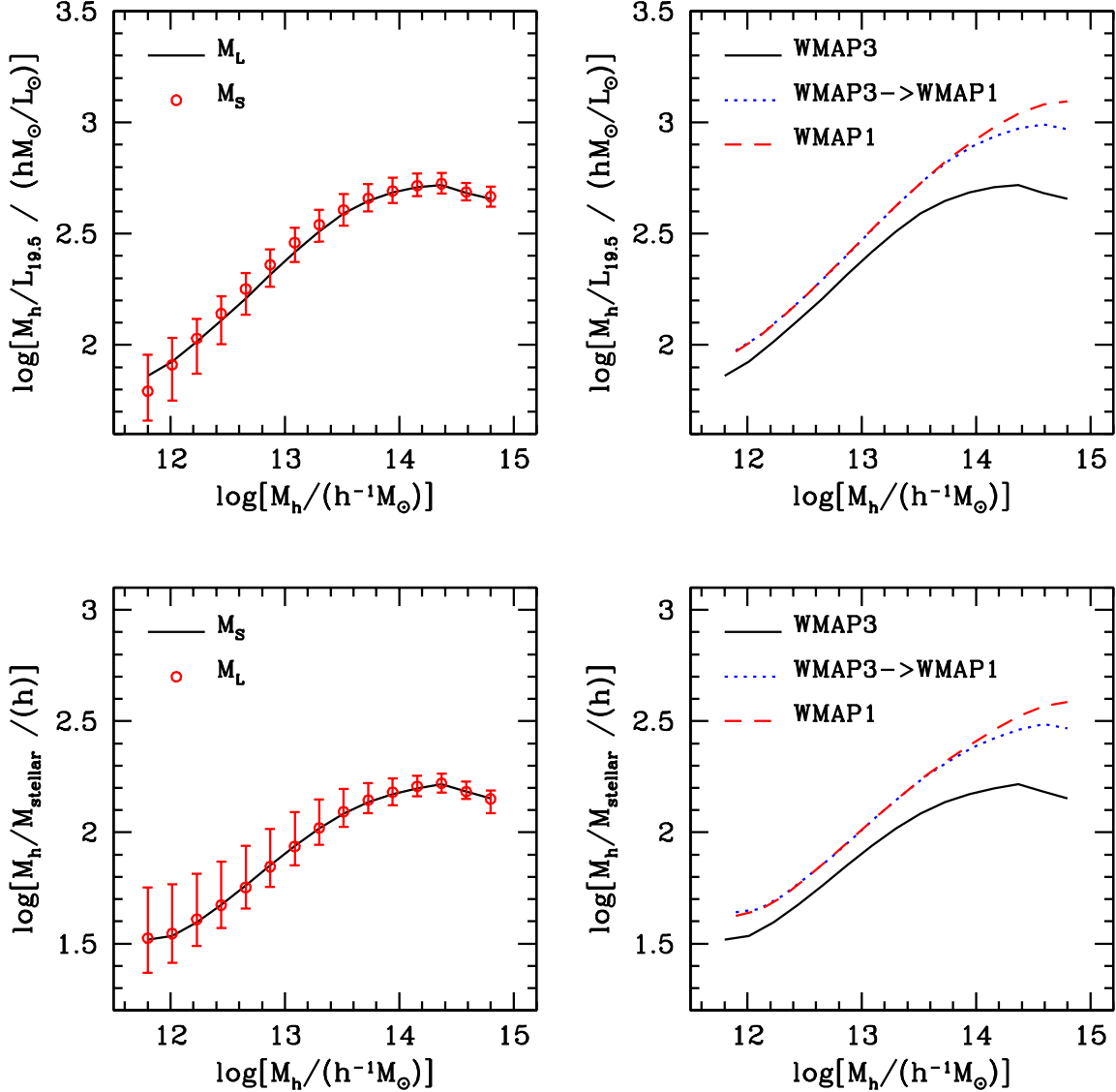}
\caption{{\it Upper left-hand panel:} the inferred mass-to-light ratios,
  $M_h/L_{-19.5}$, of  galaxy groups  in the SDSS  DR4 as  function of
  their assigned halo mass, $M_h$  (see also Table~3).  The solid line
  and  open circles  correspond to  the mass-to-light  ratios obtained
  using  the   group  masses   $M_L$  (based  on   the  characteristic
  luminosity) and  $M_S$ (based  on the characteristic  stellar mass),
  respectively.  By construction, there  is no scatter in the inferred
  mass-to-light  ratios  when using  $M_L$,  since the  characteristic
  luminosity is assumed to be related to the halo mass on a one-to-one
  basis.   {\it Lower left-hand  panel:} Same  as the  upper left-hand
  panel, except  that this  time we plot  the inferred  ratios between
  halo mass  and characteristic  stellar mass, $M_h/M_{\rm  stellar}$. 
  This time  there is no  scatter when using  $M_S$ to infer  the halo
  mass  (solid line), and  the errorbars  reflect the  68\% confidence
  levels  of  $M_h/M_{\rm  stellar}$  obtained  using  $M_L$  as  mass
  indicator. {\it Right-hand panels:} Solid  lines are the same in the
  corresponding left-hand  panels, and  are obtained assuming  a WMAP3
  cosmology throughout.   The dashed  lines show the  results obtained
  when  using a WMAP1  cosmology instead.   Finally, the  dotted lines
  correspond  to the  results  obtained when  using  the group  finder
  assuming a  WMAP3 cosmology,  but using a  WMAP1 halo  mass function
  when  inferring the  final group  masses.  See  text for  a detailed
  discussion.}
\label{fig:M_L}
\end{figure*}
   
\subsection{Average Mass-to-Light Ratios}
\label{sec:mtol}

The mass-to-light ratio of a dark matter halo expresses the efficiency
with  which stars have  formed in  that halo.   Consequently, accurate
measurements of the average mass-to-light ratios of dark matter haloes
as a function of halo mass can put tight constraints on the physics of
galaxy  formation.  The  upper left-hand  panel  of Fig.~\ref{fig:M_L}
shows  the average  mass-to-light ratios,  $M_h/L_{19.5}$,  as a
function of halo mass obtained  from our group catalogue.  Results are
shown based on both mass indicators described above: $L_{19.5}$ (solid
line), and $M_{\rm stellar}$  (open circles).  The error-bars indicate
the 68\%  percentiles from the distributions  in each $M_S$  mass bin. 
Since $M_L$ is based on the ranking of $L_{19.5}$, there is no scatter
in  the  corresponding mass-to-light  ratios.   As  one  can see,  the
mass-to-light ratios  obtained from the two  different mass indicators
are  in extremely  good  agreement  with each  other.   Note that  the
results shown  here are obtained from  Sample II: those  for Samples I
and III  are again  very similar, and  consequently not shown  for the
sake of  argument.  For  reference, all mass-to-light  ratios obtained
from Sample II are listed in Table~3. The mass-to-light ratios have 
also been obtained by various observations, e.g. 
Carlberg \etal (1996) from CNOC sample, Popesso \etal (2004) 
from RASS-SDSS. We defer a more detailed comparison to 
previous results in a forthcoming paper. 

In addition to the  mass-to-light ratios, $M_h/L_{19.5}$, we can
also use our  group catalogues to compute the  ratio between halo mass
and  characteristic stellar  mass, $M_h/M_{\rm  stellar}$.  The
lower  left-hand panel  of Fig.~\ref{fig:M_L}  shows the  average halo
mass  to  stellar  mass  ratios,  $M_{h}/M_{\rm  stellar}$,  as  a
function of  halo mass. Once again  we show the  results obtained from
both  mass indicators.   This  time the  open  circles with  errorbars
reflect the results obtained using $L_{19.5}$ as mass indicator, while
the results  based on the characteristic  stellar mass are  shown as a
solid line.  Since $M_S$ is based on the ranking of $M_{\rm stellar}$,
this  time  there   is  no  scatter  in  the   results  based  on  the
characteristic  stellar mass,  while  the errorbars  reflect the  68\%
percentiles of the distributions  in $M_{h}/M_{\rm stellar}$ where
$M_{h}$ is  obtained from  $L_{19.5}$. As  for  the mass-to-light
ratios,  the  results based  on  both  mass  indicators are  extremely
similar and are listed in Table~3.
\begin{deluxetable*}{ccccccccc}
\tabletypesize{\scriptsize} 
\tablecaption{Ratios between halo mass and luminosity and 
              between halo mass and stellar mass}

\tablewidth{0pt}
\tablehead{ 
$\log [M_{h}/\msunh]$ & $\log [M_{L}/L_{19.5}]$ &
\multicolumn{3}{c}{$\log [M_{S}/L_{19.5}]$} & $\log [M_{S}/M_{\rm stellar}]$ &
\multicolumn{3}{c}{$\log [M_{L}/M_{\rm stellar}]$} \\
\cline{3-5} \cline{7-9} \\
 & & \colhead{16\%} & \colhead{50\%} & \colhead{84\%} 
 & & \colhead{16\%} & \colhead{50\%} & \colhead{84\%}\\
\cline{1-9}\\
(1) & (2) & (3) & (4) & (5) & (6) & (7) & (8) & (9)
}

\startdata
  11.80 &   1.860 &   1.659 &   1.793 &   1.957 &   1.518 &   1.335 &   1.491 &   1.741\\
  12.00 &   1.921 &   1.736 &   1.900 &   2.025 &   1.532 &   1.411 &   1.541 &   1.763\\
  12.20 &   2.003 &   1.777 &   1.988 &   2.099 &   1.587 &   1.479 &   1.600 &   1.807\\
  12.40 &   2.089 &   1.876 &   2.100 &   2.192 &   1.660 &   1.554 &   1.660 &   1.858\\
  12.60 &   2.183 &   2.043 &   2.213 &   2.291 &   1.738 &   1.633 &   1.731 &   1.921\\
  12.80 &   2.281 &   2.215 &   2.324 &   2.394 &   1.822 &   1.723 &   1.813 &   1.992\\
  13.00 &   2.379 &   2.332 &   2.421 &   2.491 &   1.906 &   1.805 &   1.895 &   2.058\\
  13.20 &   2.470 &   2.421 &   2.504 &   2.570 &   1.983 &   1.887 &   1.973 &   2.117\\
  13.40 &   2.551 &   2.500 &   2.573 &   2.641 &   2.052 &   1.964 &   2.049 &   2.165\\
  13.60 &   2.616 &   2.562 &   2.629 &   2.694 &   2.106 &   2.052 &   2.115 &   2.205\\
  13.80 &   2.662 &   2.615 &   2.672 &   2.737 &   2.151 &   2.100 &   2.160 &   2.231\\
  14.00 &   2.693 &   2.646 &   2.699 &   2.755 &   2.180 &   2.134 &   2.187 &   2.247\\
  14.20 &   2.714 &   2.675 &   2.718 &   2.775 &   2.203 &   2.170 &   2.211 &   2.259\\
  14.40 &   2.718 &   2.681 &   2.723 &   2.767 &   2.216 &   2.178 &   2.220 &   2.261\\
  14.60 &   2.680 &   2.645 &   2.681 &   2.718 &   2.181 &   2.145 &   2.179 &   2.216\\
  14.80 &   2.657 &   2.621 &   2.667 &   2.712 &   2.152 &   2.086 &   2.151 &   2.189
\enddata

\tablecomments{Column (1): logarithm of the assigned halo mass. Column (2):
  average of the logarithm of the ratio between the assigned halo mass $M_L$
  and the characteristic luminosity $L_{19.5}$, Columns (3)-(5): 16, 50 and 84
  percentiles of the distributions of the logarithm of the ratio between the
  assigned halo mass $M_S$ and the characteristic luminosity. Column (6):
  logarithm of the ratio between assigned halo mass $M_S$ and the
  characteristic stellar mass $M_{\rm stellar}$. Columns (7)-(9): 16, 50 and
  84 percentiles of the distributions of the logarithm of the ratio between
  the assigned halo mass $M_L$ and the characteristic stellar mass. The
  mass-to-light ratios $M_{L}/L_{19.5}$ and $M_{S}/L_{19.5}$ are in unit of
  $h\Msun/\Lsun$, and the ratios between halo mass and the characteristic
  stellar mass $M_{L}/M_{\rm stellar}$ and $M_{S}/M_{\rm stellar}$ are in unit
  of $h$. All these results correspond to a WMAP3 cosmology, and we emphasize
  once more that all luminosities are in the SDSS $r$-band and have been $K+E$
  corrected to $z=0.1$}
\end{deluxetable*}

Since  our mass  assignments require  use of  the halo  mass function,
which is cosmology  dependent, it is important to  investigate how the
average $M_{h}/L_{19.5}$ and $M_{h}/M_{\rm stellar}$ change if
we change  cosmology.  The  dashed lines in  the right-hand  panels of
Fig.~\ref{fig:M_L}  show the  results obtained  when adopting  a WMAP1
cosmology   ($\Omega_m   =   0.3$,  $\Omega_{\Lambda}=0.7$,   $h=0.7$,
$\sigma_8=0.9$) instead of a  WMAP3 cosmology.  Changing the cosmology
changes (i)  the luminosity and  angular distances of all  galaxies in
the  SDSS  DR4, and  thus  their  absolute  magnitudes and  (comoving)
separations, and (ii) the halo mass function. The former has an almost
negligible (small at massive end) effect, mainly because our sample of 
galaxies is restricted
to  $z<0.2$. The  halo mass  function, however,  has a  strong impact:
Since there  are more massive  haloes in a  WMAP1 cosmology than  in a
WMAP3 cosmology,  the ranking  assigns a larger  halo mass to  a given
group,  which results  in larger  values of  $M_{h}/L_{19.5}$ and
$M_{h}/M_{\rm stellar}$.

Note that the  mass-to-light ratios are also used  in the group finder
to assign  memberships to groups (see  Section \ref{sec:steps}).  This
suggests that whenever one decides  to change one or more cosmological
parameters,  one has  to rerun  the entire  group finder  in  order to
obtain  new  mass  estimates  (as  we  did for  the  WMAP1  and  WMAP3
cosmologies shown  in Fig.~\ref{fig:M_L}). This is  impractical if one
intends  to use  the group  catalogue to  constrain  different
cosmological models. In  order to test whether we  can avoid having to
rerun the group finder when  changing cosmology we proceed as follows. 
We run our group finder over  the SDSS DR4 assuming a WMAP3 cosmology,
but then,  when we convert  $L_{19.5}$ or $M_{\rm stellar}$  into halo
mass we  use the WMAP1 halo  mass function.  The results  are shown in
the right hand panels of Fig.~\ref{fig:M_L} as dotted lines.  Clearly,
the impact  of assuming a different  cosmology in the  group finder is
almost negligible.  This demonstrates  that one can simply convert the
$M_{h}/L_{19.5}$ and $M_{h}/M_{\rm stellar}$ listed in Table 3
to another  cosmology (as  long as  it is not  too different  from the
WMAP3  cosmology adopted  here),  without having  to  rerun the  group
finder over the data, by using the relation
\begin{equation}
\label{massconvert}
\int_{M_h}^{\infty} n(M'_h) \rmd M'_h = \int_{\widetilde{M_h}}^{\infty} 
\tilde{n}(M'_h) \rmd M'_h \,.
\end{equation}
Here $M_h$ and  $n(M_h)$ are the mass and halo mass  function in the WMAP3
cosmology,   and   $\widetilde{M_h}$    and   $\tilde{n}(M_h)$   are   the
corresponding values  in the  other cosmology.  
\begin{figure*}
\plotone{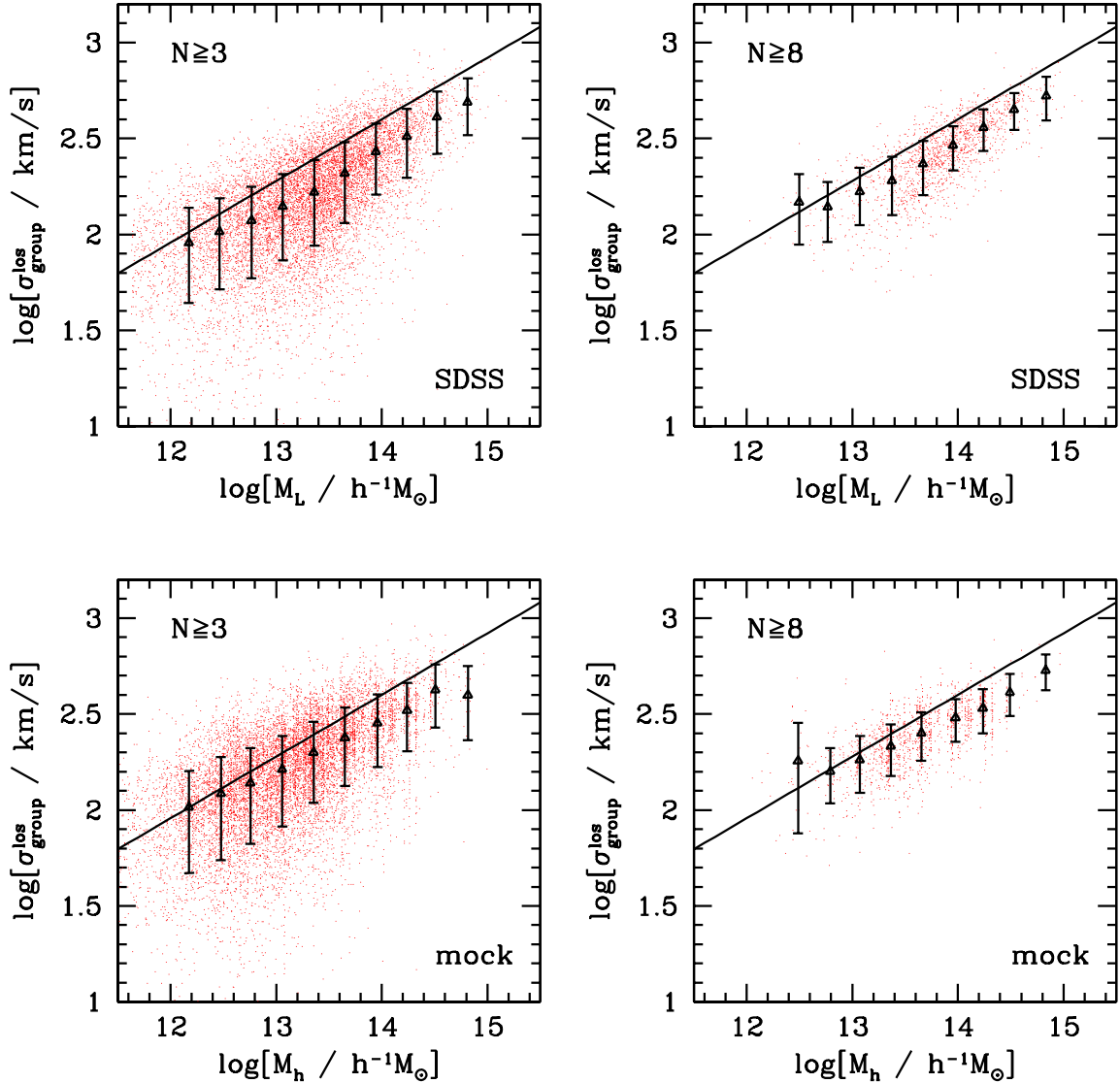}
\caption{The line-of-sight velocity dispersions  of galaxies in
  groups, obtained  using the gapper estimator of  Beers \etal (1990),
  as  function of halo mass. In the upper panels  we show  results for
  groups in our  SDSS DR4 group catalogue as  function of the assigned
  halo mass $M_L$. In the  lower panels the results have been obtained
  from the  group catalogue extracted from  our MGRS and  are shown as
  function of  the true halo  mass $M_h$. Left- and  right-hand panels
  show the results for groups with  at least 3 and at least 8 members,
  respectively. Solid  triangles with errorbars indicate  the mean and
  the  1-$\sigma$ scatter  in  each  mass bin,  while  the solid  line
  reflects    the   theoretical    expectation    values   based    on
  equation~(\ref{veldispfit}).    As  discussed   in  the   text,  the
  line-of-sight velocity dispersions of  group members are biased low,
  due to the  fact that galaxies with the  highest peculiar velocities
  in a group are the most likely to be missed by the group finder.}
\label{fig:v_M}
\end{figure*}

\subsection{Group Velocity Dispersions}
\label{sec:sigma}

For relatively massive groups, especially for groups with a sufficient
number of  member galaxies,  one can estimate  a dynamical  group mass
based  on the velocity  dispersion of  the member  galaxies. Following
Yang \etal  (2005a), we use  the gapper estimator described  by Beers,
Flynn  \&  Gebhardt  (1990)  to estimate  the  line-of-sight  velocity
dispersion of each individual group.  The method involves ordering the
set of  recession velocities  $\{v_i\}$ of the  $N$ group  members and
defining gaps as
\begin{equation}
g_i=v_{i+1}-v_i,~~~ i=1,2,...,N-1\,.
\end{equation}
The rest-frame velocity dispersion is then estimated by
\begin{equation}
\label{eq:vgap}
\sigma_{\rm gap} = \frac{\sqrt{\pi}}{(1+z_{\rm group})N(N-1)}
\sum_{i=1}^{N-1}w_i g_i\,.
\end{equation}
where the weight is defined as $w_i=i(N-i)$.  Since there is a central
galaxy in each  group, which is assumed to be at  rest with respect to
the  dark matter  halo, the  estimated velocity  dispersion has  to be
corrected. This results in a final velocity dispersion given by
\begin{equation}\label{eq_gapper}
\sigma=\sqrt{\frac{N}{N-1}} \sigma_{\rm gap}\,.
\label{eq:sigma}
\end{equation}
The upper panels of Fig.~\ref{fig:v_M} show the line-of-sight velocity
dispersions of groups thus obtained as a function of the assigned halo
mass $M_L$ for groups with at least 3 (upper left-hand panel) and with
at  least 8 members  (upper right-hand  panel).  Solid  triangles with
errorbars  indicate  the  mean  and  the  1-$\sigma$  scatter  of  the
line-of-sight velocity dispersion in each mass bin.  Clearly, there is
a good correlation between the velocity dispersion and the mass $M_L$,
indicating  that the  assigned masses  are reliable  mass  indicators. 
Compared to the theoretical prediction of equation~(\ref{veldispfit}),
which is shown as a solid line, the line-of-sight velocity dispersions
of the group  members are on average $\sim  40\%$ lower.  As discussed
in Yang \etal (2005a), this discrepancy is mainly due to the fact that
galaxies with the highest peculiar  velocities in a group are the most
likely to  be missed  by the group  finder.  To demonstrate  this, the
lower  panels  of Fig.~\ref{fig:v_M}  show  the corresponding  results
obtained  from our  mock group  catalogue.   In this  case, the  input
velocity dispersions for galaxies in  haloes have a mean relation that
is given  by the solid  line, and the  halo masses $M_h$ are  the true
halo masses.  Here  again, we see that the  velocity dispersions among
the selected  group members  are lower than  that implied by  the halo
masses.

\section{Summary}
\label{sec_conclusion}

In this paper, we have used a modified version of the halo-based group
finding algorithm  developed in Yang \etal (2005a)  to construct group
catalogues from the  SDSS DR4.  Changes and improvements  in the group
finder have been made in the following aspects:
\begin{itemize}
\item In order to assign group memberships, we need to estimate masses
  for  all  tentative groups.   Rather  than  using  a model  for  the
  mass-to-light   ratios,   as  we   did   previously,   we  now   use
  self-consistent  mass-to-light   ratios  obtained  from   the  group
  catalogue  in  an  iterative  way.
\item In  order to correct  the characteristic luminosity  and stellar
  mass for missing  members due to the magnitude  limit of the survey,
  we use the mean correction factors obtained  self-consistently  from 
  groups at low redshifts.
\item We have corrected the survey edge effect on the groups by a 
  correction factor.
\item In  order to  estimate group masses,  we use two  different mass
  indicators, one based  on the characteristic luminosity, $L_{19.5}$,
  and the other on the characteristic stellar mass, $M_{\rm stellar}$.
\end{itemize}
Tests based on detailed mock SDSS DR4 catalogues show that $\sim 80\%$
of all  groups have  a completeness $>  80\%$. The fraction  of groups
with a  completeness of $100\%$ ranges  from $\sim 60\%$  for the most
massive  groups  to $>  95\%$  for groups  with  masses  in the  range
$10^{12.5}\msunh < M_h < 10^{13} \msunh$.  On the order of $85\%$ of all
groups have an interloper fraction  $< 50\%$, while $\sim 65\%$ of the
groups have zero interlopers.

We have applied  our group finder to three  galaxy samples constructed
from  the SDSS  DR4 galaxy  catalogue: Sample  I, which  only contains
galaxies with measured redshifts from  the SDSS; Sample II, which also
contains those  SDSS galaxies for  which redshifts are  available from
alternative sources  (mainly from the  2dFGRS); and Sample  III, which
also includes  galaxies which  due to fiber  collisions do not  have a
measured redshift, but which have  been assigned the redshift of their
nearest  neighbor.   We  obtain  a  total of   $295992$,  $301237$ and
$300049$ groups  from Samples  I, II and  III, respectively,  and each
group is assigned two values for its halo mass based on the ranking of
either  the characteristic  luminosity or  the  characteristic stellar
mass of its member galaxies. 

In this  paper we have presented  some of the basic  properties of the
group catalogue,  such as the distributions of  richness, redshift and
mass.  In addition  we have presented the average  ratios between halo
mass  and   characteristic  luminosity  and  between   halo  mass  and
characteristic stellar mass.   Although these are cosmology dependent,
we have  demonstrated that it  is straightforward to convert  these to
other cosmologies.   A more detailed analysis of  the group properties
and  their   implications  for  halo   occupation  statistics,  galaxy
formation and cosmology  will be presented in a  series of forthcoming
papers.  As  a  final  note,  we mention  that  all  group  catalogues
presented here are available from the authors upon request.


\section*{Acknowledgments}

We thank Daniel H. McIntosh  for providing us the fitting function for
the  stellar mass-to-light  ratio  as  a function  of  color for  SDSS
galaxies, and the anonymous referee for helpful comments that improved
the presentation of this paper. FvdB acknowledges exhilarating 
discussions with Alison Coil
on the topic of edge-effects.  XY is supported by the {\it One Hundred
  Talents}  project and the  Knowledge Innovation  Program (Grant  No. 
KJCX2-YW-T05) of the Chinese Academy of Sciences, and grants from NSFC
(Nos.10533030, 10673023).   HJM would like to  acknowledge the support
of  NSF  AST-0607535,  NASA   AISR-126270  and  NSF  IIS-0611948.   MB
acknowledges  financial  support  from  the  Austria  Science  Council
through Grant P18416.

Funding for  the SDSS and SDSS-II has  been provided by the  Alfred P.
Sloan Foundation, the Participating Institutions, the National Science
Foundation, the  U.S.  Department of Energy,  the National Aeronautics
and Space Administration, the  Japanese Monbukagakusho, the Max Planck
Society, and  the Higher Education  Funding Council for  England.  The
SDSS Web  Site is  http://www.sdss.org/.  The SDSS  is managed  by the
Astrophysical Research Consortium  for the Participating Institutions.
The  Participating Institutions  are  the American  Museum of  Natural
History,  Astrophysical   Institute  Potsdam,  University   of  Basel,
Cambridge University,  Case Western Reserve  University, University of
Chicago,  Drexel  University,  Fermilab,  the Institute  for  Advanced
Study, the  Japan Participation  Group, Johns Hopkins  University, the
Joint  Institute for  Nuclear  Astrophysics, the  Kavli Institute  for
Particle Astrophysics  and Cosmology, the Korean  Scientist Group, the
Chinese Academy of Sciences  (LAMOST), Los Alamos National Laboratory,
the     Max-Planck-Institute     for     Astronomy     (MPIA),     the
Max-Planck-Institute   for  Astrophysics   (MPA),  New   Mexico  State
University,   Ohio  State   University,   University  of   Pittsburgh,
University  of  Portsmouth, Princeton  University,  the United  States
Naval Observatory, and the University of Washington.


\label{lastpage}

\end{document}